\documentstyle[12pt,epsf]{article}
\begin{document}


\def\be{\begin{equation}}
\def\bea{\begin{eqnarray}}
\def\ee{\end{equation}}
\def\eea{\end{eqnarray}}
\def\su{\subset}
\def\ri{\rightarrow}
\def\tilde{\widetilde}


\def\P{\Pscr}
\def\C{\Cscr}
\def\p{\phi}
\def\n{\nu}
\def\r{\rho}
\def\e{\varepsilon}
\def\d{\delta}
\def\D{\Delta}
\def\th{\teta}
\def\a{\alpha}
\def\t{\tau}
\def\g{\gamma}
\def\s{\sigma}
\def\l{\lambda}
\def\O{\Omega}
\def\o{\omega}
\def\b{\beta}
\def\m{\mu}

\textwidth 400pt
\topmargin  5 mm
\oddsidemargin 8mm
\evensidemargin 8mm
\baselineskip 5mm
\textheight 580pt
\renewcommand{\baselinestretch}{1.3} 
\hyphenation{pro-pa-ga-tion}

\title{Closing probabilities in the Kauffman model: an annealed computation.}
\author{U. Bastolla and G. Parisi}
\maketitle
\centerline{Department of Physics, University ``La Sapienza'', P.le Aldo
Moro 2, I-00185 Roma, Italy}
\date{}
\medskip
\centerline{Keywords: Disordered Systems, Genetic Regulatory Networks,}
\centerline{Random Boolean Networks, Cellular Automata}
\begin{abstract}

We define a probabilistic scheme to compute the distributions of periods,
transients and weigths of attraction basins
in Kauffman networks. These quantities are obtained in the
framework of the annealed approximation, first introduced by Derrida and
Pomeau. Numerical results are in good agreement with
the computed values of the exponents of average periods, but show also
some interesting features which can not be explained whithin the annealed
approximation.

\end{abstract}
\newpage
\section{Introduction}

Kauffman Networks are a disordered dynamical system. They were introduced in
1969 as a simplified model of the genetic regulatory system acting in
cell differentiation \cite{K69}. The debate about cellular automata grew
the interest of physicists about the model \cite{K84}. It was soon
recognized \cite{DF1} that it has deep analogies with a disordered system
studied in statistical mechanics, the infinite range Spin Glass, whose
properties had been investigated and understood in the immediately former
years \cite{MPV}.

A configuration of the system consists of $N$ binary variables, $\s_i=0,1$,
$i=1\cdots N$. In the biological metaphor, the variable $\s_i$
represents the state of
activation of the $i$-th gene in the cell.

The configuration space is provided of a metric structure through the
Hamming distance (normalized to one), say the fraction of
variables whose states are different between two configurations $C$ and $C'$:
\be d\left(C,C'\right)
={1\over N}\sum_{i=1}^N\left({\s_i-\s'_i\over 2}\right)^2. \ee

The system is disordered in the
sense that the dynamic is deterministic, but its rules are chosen at random
at the beginning in the following way: for each gene $i$, $K$ genes are
extracted randomly
to send a signal to it (the same gene can be extracted more than
once); these are denoted by $j_l(i)$, $l=1\cdots K$;
for each gene $i$ and each of the $2^K$ configurations of the signal
it can receive, the value of the response function $f_i(\s_1\cdots\s_K)$
is extracted to be 0 with probability $p$ and 1 with probability $1-p$.
The dynamic rules are then quenched and do not change during the evolution
of the system.

At each time step the $N$ variables are updated simultaneously, according to
the signal received and to the response to that signal:

\be \s_i(t+1)=f_i\left(\s_{j_1(i)},\cdots\s_{j_K(i)}\right). \ee

In a more compact way, the evolution law can be put in the form $C(t+1)=
\Phi_\eta\left(C(t)\right)$, where $C$ denotes a configuration of the dynamic
variables and $\eta$ a configuration of the dynamic rules.

Due to the fact that configuration space is finite, every trajectory ends up,
after a transient time, on a periodic orbit, and configuration space is
partitioned into a certain number of attraction basins of weight $W_\a$,
where the $\a$ labels the periodic orbit.
So, as it is customary in the theory of disordered systems, three kinds of
average have to be defined:
\begin{itemize}\item{The time average, which is equivalent to the
average over a given limit cycle.}
\item{The average over the same network, where
different cycles have to be weighted with their own $W_\a$.}
\item{The average over different realizations of the network.}
\end{itemize}

It was soon discovered that Kauffman Networks have two different kinds of
dynamical behaviours: a $frozen$ regime, when the average cycle length
grows as a power of $N$, and a $chaotic$ regime when it grows exponentially
with $N$.

It is possible to characterize this transition as a phase
transition and two order parameters have been found to describe it.

Derrida and Pomeau considered the average limit distance between
configurations on two randomly chosen trajectories in the same network.
They computed its value in the framework of the annealed approximation
and found that it is zero in the frozen phase (in the limit of an infinite
network: remember that the distance is defined as the fraction of different
variables), while it is different from zero in the chaotic one \cite{DP,HN,DS}.

Then Flyvbjerg introduced the concept of {\it stable core}, defined as the set
of the variables that evolve to a constant state not depending on the initial
configuration. The fraction of variables belonging to the stable core
is 1 (in the limit of infinite $N$) in the frozen phase while is less
than 1 in the chaotic phase \cite{F1}.

The transitions described by these parameters take place at the same critical
point, $p=p_c(K)$, where $p_c$ is such that $2p_c(1-p_c)=1/K$
\cite{DP,HN,DS,F1}.

Despite of the success of this approach, the connection between these
order parameters and dynamical properties (of the kind of the average cycle
length or the weights of attraction basins) is still lacking.
This work is a contribute to fill in this gap.

In the second section we define the closing probabilities, which are the
quantities that allow to compute the full cycle length and transient time
distribution.
Section 3 describes the annealed approximation, first introduced by
Derrida and Pomeau \cite{DP}, that here is used to compute the distribution
of the overlap between different time configurations and the closing
probabilities, whence we obtain the distribuition of cycle lengths (section 4)
and the distribution of attraction basins weigths in the chaotic phase
(section 5), which turn out to be the same as those obtained by Derrida and
Flyvbjerg in the limit case $K\rightarrow\infty$ (the so called Random Map)
\cite{DF2}.
Section 6 deals with the average number of cycles in a given network, which
is computed in the chaotic phase, in the framework of the annealed
approximation.

In section 7 numerical results  are presented. This section is subdivided into
three subsections: in the first one, data concerning the distribution
of the overlap are shown; they support the validity of the annealed
approximation, but only in some range of the temporal variables. In section
$7.2$ simulations about the closing probabilities are reported, which show
deviations from the annealed approximation after an initial agreement.
In section $7.3$ we show the exponents which give the $N$ behaviour of the
average cycle length for different values of the parameter $K$; there is
a good agreement between our computations and the numerical results, but
the distribution of cycle lengths that we measured is quite different from the
predicted one.

In the last section we interpret the meaning of the approximation used and
the discrepancies that we noticed with the simulations, and point out
the direction of future work.

\section{Closing probabilities}

In the following, we will use the overlap $q(C,C')=1-d(C,C')$ to describe
the metric of configuration space.

We will follow a stochastic strategy, with the aim to compute the probability
distribution of the overlap $q(t,t')$ between the configurations $C(t)$ and
$C(t')$, $t'>t$, on the same trajectory. It is convenient to consider only
trajectories not yet closed at time $t'-1$, so what we actually want to compute
is a conditional probability, that we denote by the symbol $P_N(t,t';q)$.

We define the closing probability $\pi_N(t,t')$ as the probability to find
$q(t,t')=1$, which means that the two configurations are equal,
over a trajectory not yet closed at time $t'-1$:

\be \pi_N(t,t')=P \left\{ q\left(t,t'\right)=1\mid q(t_1,t_2)< 1,
t_1<t_2< t'\right\}.  \ee

We will have also to deal with the integral closing
probability, $\tilde\pi_N(t)$, which is the probability that a trajectory
not yet closed closes at time $t$ regardless of the length of the cycle
reached, $\tilde\pi_N(t)=\sum_l\pi_N(t-l,t)$.

Then it is easy to compute $F_N(t)$, the probability that a
trajectory has not yet closed at time $t$. It satisfies the equation
$F_N(t+1)=F_N(t)\left(1-\tilde\pi_N(t)\right)$,
whence, in the limit of large systems, introducing a continuous time variable,
and using the fact that $\tilde\pi_N(t)$ is small even at times when nearly
every trajectory has closed, one finds the simple expression

\be \label{F}
F_N(t)=\exp\left( -\int_0^t \tilde\pi_N(\t){\rm d}\t\right), \ee
so that the probability to find a trajectory that, after a transient time
$t$, reaches a cycle of length $l$, is given by

\be \label{periodo}
P\left\{ T=t,L=l\right\} =\exp\left( -\int_0^{t+l} \tilde\pi_N(\t){\rm d}\t
\right) \pi_N (t,t+l). \ee

\section{Annealed approximation: the overlap distribution}

The goal to compute the overlap distribution becomes easier if one can reduce
the dynamics of the overlap to a stochastic process.

We can treat Kauffman model
as a stochastic process by extracting the values of the response functions not
at the beginning, but every time a variable receives a signal that it had
never received, and using the values already extracted every time the same
signal comes back. Note that, if every gene receives signals from the whole
system, we have not to worry about this last eventuality, which never occurs
before the trajectory closes. In this case the annealed approximation that we
are going to define is exact. This case corresponds to the limit of infinite
$K$
and is known as the Random Map model, studied ananlitically by Derrida and
Flyvbjerg  \cite{DF2}. In this work, we generalize their results to finite
$K$ values, in the framework of the annealed approximation.
\vspace{0.5 cm}

Thus, in order to compute the overlap between two configurations, for every
gene
we distinguish two cases: either the gene received the same signal $S_i(t)$ at
the two previous time steps, and so it will be in the same state, or it didn't:

\be q(t+1,t'+1)={1\over N}\sum_i \d_{S_i(t)S_i(t')}
+{1\over N}\sum_i\left( 1-\d_{S_i(t)S_i(t')}\right)\d_{f\left(S_i(t)\right)
f\left(S_i(t')\right)}. \ee

Now, the annealed approximation consists in this: we extract at random
{\it at every time step} all the connections in the network, keeping memory
only of the minimal information about the overlap $q(t,t')$, so that the
probability that the two signals $S_i(t)$ and $S_i(t')$ are the same is
$q(t,t')^K$; if they are not equal, we extract at random the values of the
response function, and they will be equal with probability $1-\rho$, where
$\r=2p(1-p)$ \cite{DP}.

Derrida and Pomeau introduced the annealed approximation in order to obtain
the equations for the evolution of the average Hamming distance between two
randomly chosen configurations; we will use it in order to treat the overlap
as a Markovian stochastic process, and so to compute the closing probabiliies.
We will argue later from our simulations that the overlap distribution
obtained through the annealed approximation is very close to the quenched one.

It is not possible, using this kind of approximation, to impose the condiction
that the trajectory where we measure $q(t,t')$ was not closed at time $t'-1$:
we can only impose that it did not close on a cycle of length $l=t'-t$.
As we shall see, this fact is likely to be the source of the main discrepancies
between the annealed approximation and our simulations.

If we impose this condiction, we obtain the master equation

\be \label{master} P_N\left(t+1,t+l+1;q_n\right)=
{N\choose n}
{\sum_{m=0}^{N-1} P_N\left(t,t+l;q_m\right) (\g (q_m))^n
(1-\g (q_m))^{N-n} \over 1- P_N \left(t,t+l;1\right) }, \ee
where $q_m=m/N$ and

\be \label{gamma}\g (x)=(1-\r)+\r x^K. \ee

Given an initial distribution $P_N(0,l; q)$, the process defined by
(\ref{master}), which is ergodic because we exclude the overlap $q=1$,
evolves to a stationary distribution independent on both $t$ and $l$,
which appears only in the initial distribution but not in the transition
probability.

We look for a stationary distribution $\tilde P_N$ in an exponential form,
\be \tilde P_N (q=x)=C_N (x){\rm e}^{-N\a (x)}. \ee

In the limit $N\rightarrow\infty$ we can use continuous variables, transform
the sum into an integral and apply the saddle point method to perform it,
obtaining the following selfconsistence equations for the unknown functions
$\a(x)$ and $q(x)$ (the last one is the saddle point of the integral for a
given value of $x$):
\bea\label{alpha}
& \a^\prime(q(x))-\g^\prime(q(x))\hspace{0.2 cm}\left( {x\over \g(q(x))}-
{1-x\over 1-\g(q(x))}\right)=0  \\
\vspace{0.5 cm}
& \a(x) =\a(q(x))+x\log\left({x\over \g(q(x))}\right)
+(1-x)\log\left({1-x\over 1-\g(q(x))}\right), \label{alpha2}
\eea
with the condition $q(x)<1$.

Deriving the first equation and inserting the result in the second one we get a
non-local transcendent equation for the function $q(x)$:

\be \g'\left(q(x)\right)\left({x\over \g\left(q(x)\right)}-
{1-x\over 1-\g\left(q(x)\right)}\right)=\log\left({q(x)\over 1-q(x)}\right)
-\log\left({\g\left(q\left(q(x)\right)\right)\over
1-\g\left(q\left(q(x)\right)\right)}\right). \label{q(x)}\ee

We were not able to solve this equation, so we had to solve numerically a
discrete version of it in order to obtain the exponent of the closing
probability, $\a(1)$.

Before to turn to this point, however, there are some important features
of the overlap distribution that can be easily understood.


Deriving (\ref{alpha2}), we can see that the point $Q(t)$ where
the overlap distribution is concentrated satisfies the equation

\be Q(t+1)=\g\left(Q(t)\right)\label{q2}, \ee
which, at stationarity, becomes

\be Q^*=\g (Q^*)=1-\r +\r{Q^*}^K. \label{Q*}\ee

This equation was first derived by Derrida and Pomeau as a mean field
solution of the annealed model \cite{DP}.
The condition $\g'(1)=K\r =1$ separates parameter space in two different
regions:
\begin{itemize}
\item{$K\r\le 1$: {\it frozen phase}. The only fixed point of the map
(\ref{q2})
is $Q^*=1$. This implies, as we shall see, that $\a(1)$ goes to zero and
the typical cycle length increases less than exponentially with system size.}
\item{$K\r >1$: {\it chaotic phase}. The fixed point $Q^*=1$ becomes unstable
(moreover, it would give raise to a negative variance) and the overlap is
concentrated around another fixed point of (\ref{q2}), $Q^*<1$. In this case
$\a(1)$ is finite and the typical cycle length increases exponentially with
system size.}
\end{itemize}

The variance of the distribution can also be easily obtained, in the
Gaussian approximation, deriving (\ref{alpha2}) twice and imposing the
saddle point condition. One finds that the variance multiplied by $N$,
$V(t)$, follows the recursive equation

\be V(t+1)=(\g^\prime\left(Q(t)\right)^2 V(t)+Q(t+1)(1-Q(t+1)), \ee
whose fixed point is
\be
V^*={Q^* (1-Q^*)\over 1-(\g^\prime(Q^*))^2} \ee

It is appearent that the stationary overlap distribution is a non binomial one:
the presence of correlations between the genes are made evident by the fact
that $V^*$ is larger than $Q^*(1-Q^*)$. Moreover, it is also easy to see that
$\a'(1)$ is approximately equal to $\r K$ and does not diverge, while in the
binomial distribution it does.

On the other hand, when $K$ tends to infinity destroing the correlations
between consecutive overlaps, the stationary overlap distribution becomes a
binomial one and $Q^*$ tends exponentially to $1-\r$.

In the frozen phase, when $\g'(1)\leq 1$, $Q^*=1$ is the true solution of
equation (\ref{Q*}), and $\a(1)$ vanishes or, to say better, is of order $1/N$.

It is important to see how the term of $O(1)$ vanishes with time: to see that,
let us see what happens at the critical point $\r=1/K$.

The evolution of the most probable overlap at the critical point, for $K>1$,
is such that, introducing a continuous time variable,
\be
Q^\prime (t)={1\over K}(Q-1)^2 \sum_{i=1}^{K-1} (K-1-i)Q^i,\ee
and the asymptotic behaviour in $t$ is
\be 1-Q\propto \left({(K-1)t\over 2}\right)^{-1}. \ee
\vspace{0.5 cm}

Let's now turn to the numerical computation of the exponent of the closing
probability, $\a(1)$.

We have to use the discrete variables $q_i=i/N$; moreover, in place of the
stationary state equation, we consider the evolution of the exponent of the
overlap distribution, $\a_t(\{q_i\})$ (we omit to specify the dependence on
$l$, which appears only in the initial condition $\a_0(\{q_i\})$).

Neglecting the normalization factor, in the master equation, we get the
following map for the evolution of $\a_t(q_i)$:

\bea
& \a_{t+1}\left(q_i\right)=q_i\log\left(q_i\right)+
\left(1-q_i\right)\log\left(1-q_i\right)+\hspace{2 cm}\nonumber\\
& \hspace{2 cm}\min_{j<N}\left(\a_t(q_j)
-\left(q_i\right)\log\left(\gamma(q_j)\right)
-\left(1-q_i\right)\log\left(1-\gamma(q_j)\right)\right), \label{map}\eea

This map does not have a fixed point $\{\bar{\a}(q_i)\}$, as it can be seen
numerically: for istance, $\a_t(1)$ reaches a minimum at a certain time and
then increases uniformly.

This increase is not in contradiction with the existence of a stationary
distribution; in fact equation (\ref{map}) misses the normalization factor
$\left(1-P(1)\right)^{-1}$, and so the probability computed by means of it
has to decrease when the exact probability has reached the stationary value.

The increment is of order $O\left({1\over N^2}\right)$, because,
after $Q_N(t)$ reaches its stationary value $Q^*$, $\a(Q^*_N)$ grows of a term
of that order due to the difference, $O\left({1\over N}\right)$, between
$Q^*_N$ and $\g(Q^*_N)$, so it is not important because we also miss terms of
order ${1\over N}$ in the saddle point approximation.
In our algorithm, we decided to get the function $\a_t(q_i)$ at the
time when $\a_t(1)$ reaches its minimum value.

There is another finite $N$ effect in our numerical calculation: because of the
condition $j<N$ (this means that the overlap $x_j=1$ cannot be taken as a
starting point), the stationary value of $\a_N(x_i)$ decreases with $N$. In
fact, the larger is $N$, the lower can be the minimum in (\ref{map}), and the
effect, of order $1/N$, is important especially for $i=N$, because in this case
in the stationary state the minimum in (\ref{map}) would be attained for $j=N$.

The finite size effects become more relevant at the critical line $\rho=1/K$.
As a discretization effect, the most likely value of the annealed distribution
is not $Q^*=1$, because the corresponding value $j=N$ in the equation
(\ref{map}) is not allowed, and at stationarity $\a(x_i)$ has a minimum for
$Q^*_N=1-\epsilon_N$, where $\epsilon_N$ is $O\left({1\over\sqrt N}\right)$,
so that $\g(Q^*_N)-Q^*_N$ is of order $O\left({1\over N}\right)$.

Actually, in the frozen phase the equation (\ref{map}) is not more a good
approximation, because then we are not allowed to neglect $P(1)$ in the
denominator of equation (\ref{master}). We keep it in the complete form and
obtain the equation for the first moments of the distribution:
\be (1-\r)+\r\langle q^K\rangle=\langle q\rangle \left(1-P(1)\right)
+\left(P(1)\right), \ee
whence, at the critical point, neglecting terms of order $\langle\left( q-
\langle q\rangle\right)^3\rangle$ and $\left(1-\langle q\rangle\right)^3$,
one gets
\be P={K-1\over 2}\left(1-\langle q\rangle +{V\over N\left(1-\langle q
\rangle\right)}\right).
\label{p2}\ee

This suggests that both $P(1)$ and $1-\langle q\rangle$ are of order
$1/\sqrt N$. Combining this fact with the previous results, one finds in the
critical line an asymptotic closing probability of the form

\be \label{pk2}
\pi_N(t,t')\propto{1\over\sqrt N}\exp\left({-4N\over (K-1)t^2}\right),\ee

while in the chaotic phase we find an exponential behaviour:
\be
\pi_N(t,t')\propto {\rm e}^{\a(1)N}, \ee
provided that the smallest time $t$ is large enough for the overlap to reach
the stationary distribution.

\section{Period distribution within the Annealed Approximation}
We now use the closing probabilities obtained in the previous section
to compute the  asymptotic distribution of
transient times $T$ and cycle lengths $L$.

In the chaotic phase, inserting the expression for the asymptotic closing
probability into the equation (\ref{periodo}), we obtain

\be {\rm Prob}\left\{T=t, L=l\right\}
\simeq\exp\left(-{1\over 2}(t+l)^2 {\rm e}^{-\a N}\right)
{\rm e}^{-\a N}, \label{perann}\ee
(from now on, we will shortcut $\a(1)$ with $\a$).

So, for finite $K$, we obtain the same form of the distribution as in the
Random Map Model, differing only in the value of $\a$.

Introducing the continuous variables $x=t/\t$ and $y=l/\t$, with

\be
\label{tau}
\t={\rm e}^{{1\over 2}\a N}, \ee
the period and transient distribution becomes a continuous distribution with
density

\be f(x,y)={\rm e}^{-{1\over 2}(x+y)^2}. \label{densa}\ee

While the form of the distribution predicted by the annealed approximation is
not confirmed by our simulations, the $N$ behaviour of the time scale $\t$ is
in good agreement with the results of the simulations,  as it is shown by
figure \ref{fig_alp} where we compare the value of $\a$ obtained by the
iteration of equation (\ref{map}) with twice the exponent of the average cycle
length and transient time or with the exponent of the variance of the period
obtained by a fit of experimental data, for some $K$ value.
Other numerical results will be presented later.
\vspace{0.5 cm}

In the frozen phase $\a$ vanishes, and the average cycle length grows as a
power of $N$. The closing probability, for $K\ne 1$, is given by
\be \pi_N(t,t+l)\propto
{1\over (K-1)\t}\exp\left(-\left({2\t\over t}\right)^2\right), \ee
where the time-scale at the critical point is given by
\be \t=\sqrt{ N/(K-1)}.\ee

This suggests a kind of universality of the critical line in the Kauffman
model:
the time scale of the dynamics diverges always as the square root of the
number of the elements (except for the case $K=1$, where the variance in the
Gaussian approximation diverges linearly with $t$, thus destroying the validity
of the approximation, and our method fails).

It is easy to see that the exponent $\a$ of the closing probability vanishes
quadratically when $\r$ approaches to the critical point
(except for the case $K=1$). In fact, putting
$\r=\r_c+\epsilon$, one finds
\be Q^*=1-{2K\over K-1}\epsilon +{\rm o}(\epsilon),\ee
\be \a\approx {1\over 2}{\left(1-Q^*\right)^2\over V}={(2K)^2\over K-1}
\epsilon^2+{\rm o}\left(\epsilon^2\right)\ee
(of course these expressions are not valid in the limit case of infinite
$K$, where the overlap distribution is a binomial one with mean value $1-\r$
and $\a$ is equal to $-\log(1-\r)$).

\section{Distribution of weights}
In Kauffman networks configuration space breaks into a number of limit cycles
with their own attraction basins. The statistical distribution of the
weights of such basins, $W_\a$ (the rate of the number of configurations in the
basin to the total number), has been computed in the limit cases $K=1$
and $K=\infty$ by Derrida and Flyvbjerg \cite{DF2} and Flyvbjerg and Kjaer
\cite{FK} respectively.

In the framework of the annealed approximation it is possible to obtain this
distribution for every value of the parameters, at least in the chaotic phase:
in this case, it turns out to be the same distribution computed in \cite{DF2}
for the Random Map.

Following \cite{DF2}, it is convenient to start computing the quantities
\be \label{Yp}
\overline{Y_p}=\sum_\a\overline{W_\a^p},\ee
that are the moments (of order $p-1$) of the probability to find a basin of
weight $W$. They can be computed by noting that they represent the probability
that $p$ configurations chosen at random ultimately meet.


In the framework of the annealed approximation, one finds that these quantities
do not depend, in the chaotic phase and in the limit $N\rightarrow\infty$,
on the parameter $\t$ which gives the cycle length time scale, and have the
same values computed by Derrida and Flyvbjerg for the case of the Random Map.

To be more concrete, let's calculate $\bar Y_2$. We must generate randomly
two trajectories on the same network. We extract the first one and
follow it till the time
$T_1$ when it closes on itself, as we have done in the previous section; then
we generate another "replic" of the system and we study the overlap
$q_{12}(t_1,t_2)$ between
the  configuration at time $t_1$ on the first trajectory and the one at time
$t_2$ on the second one.

We define the closing probability of the second trajectory on the
first one as the probability that $q_{12}(t_1,t_2)$ is equal to 1,
with the condition that the first
trajectory is composed of $T_1$ different configurations,
the second of $t_2-1$ different ones and none of them has met the first
trajectory before time $t_2$ (we express this condition by the symbol
$O(T_1,t_2)$):

\be \pi_N^{(2)}(T_1;t_1,t_2)=P\left\{q_{12}(t_1,t_2)=1|O(T_1,t_2)
\right\}. \ee

The probability that the second trajectory meets the first one after
$T_2$ time steps is then
\be F_N^{(2)}(T_1,T_2)\tilde\pi_N^{(2)}(T_1;T_2), \ee
where $\tilde\pi_N^{(2)}(T_1;t_2)$ is the integral closing probability, say the
sum over $t_1$ of $\pi_N^{(2)}(T_1;t_1,t_2)$, and $F_N^{(2)}(T_1,T_2)$ is the
probability
that the second trajectory is composed of $T_2$ different configurations
and they don't touch the first trajectory before time $T_2$. One has, in the
limit of continuous time variables,

\be F_N^{(2)}(T_1,T_2)=\exp\left(-\int_1^{T_2-1}\left(\tilde\pi_N^{(2)}(T_1;t)+
\tilde\pi_N^{(2)}(t)\right)dt\right), \ee
and

\be \label{Y2}\overline{Y}=\sum_{T_1,T_2} F_N(T_1)F_N^{(2)}(T_2,T_1)
\tilde\pi_N(T_1)\tilde\pi_N^{(2)}(T_2;T_1), \ee
where $\tilde\pi_N(t)$ and $F_N(t)$ are respectively the integral closing
probability and the opening probability for only one trajectory, as
defined in the second section.

In the framework of the annealed approximation, we can imagine that the
overlap $q_{12}(t,0)$ between the initial configuration of the second
trajectory and a given configuration of the first one is a binomial variable,
and that the distribution of $q_{12}(t+1,1)$ is obtained from it through
the annealed evolution equation.

In this hypothesis, asymptotically in $t_1$ and
$t_2$, the closing probability in the chaotic phase
is given by $\pi^{(2)}_N(T_1;t_1,t_2)=1/\t^2$, where $\t=\exp(\a N/2)$ is the
cycle length time scale.

Operating the substitutions $x=T_1/\t$, $y=T_2/\t$ and transforming to
continuous variables, one then finds, for large values of $\t$,

\be \bar Y_2\simeq\int_0^\infty\int_0^\infty x^2{\rm e}^{-{(x+y)^2\over 2}}
dx dy=2/3, \ee
which is independent on $\t$ and is the same result of \cite{DF2}.

This computation generalizes easily to an arbitrary number of trajectories,
and in this way one finds for everyone of the $Y_p$ the same value obtained
by Derrida and Flyvbjerg for the Random Map; so one can argue that
$f(W)$, which has the meaning that
$f(W)dW$ is the average number of cycles with weight between $W$ and
$W+dW$, is given, in the chaotic phase and in the framework of the annealed
approximation, by the same expression valid in the Random Map model:

\be f(W)={1\over 2}W^{-1}\left(1-W\right)^{-1/2}.\label{f(W)} \ee

To compare this prediction with numerical data, we have pursued the
simulations performed by Derrida and Flyvbjerg \cite{DF1}
relatively to $\bar Y_2$ for $K=3$, doubling the size of the maximum system
simulated. For each system size we generated 10000
sample networks and two configurations at random on each of them and we
measured $Y_2$ as the probability that they end up on the same cycle, as it
was done in \cite{DF1}.

\begin{figure}
\centering
\epsfysize=15.0cm
\epsfxsize=10.0cm
\epsffile{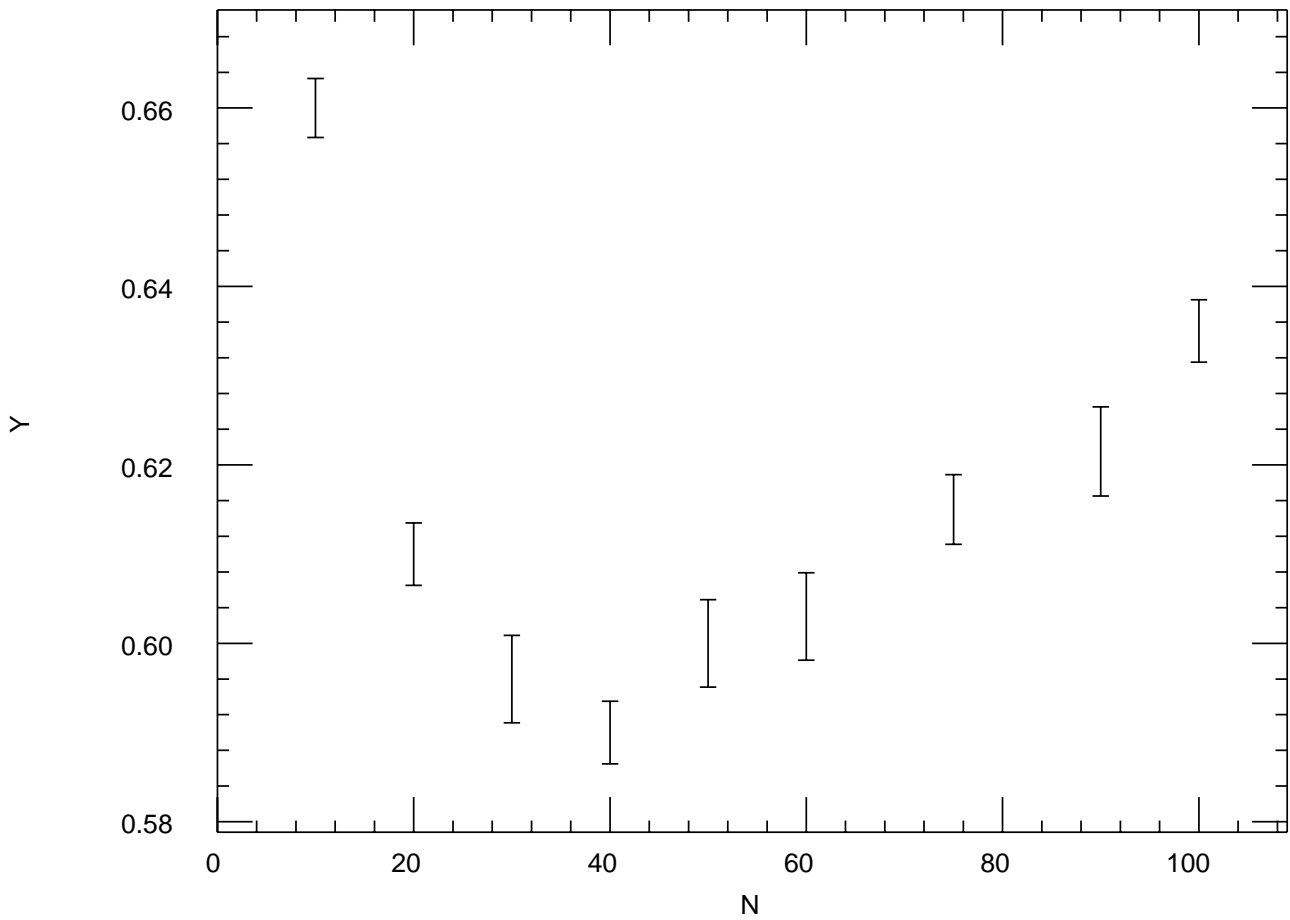}
\caption{\it $\bar{Y}_2$ versus $N$ in networks with $K=3$.The annealed
prediction is $\bar{Y}_2=0.\bar{6}$.}
\label{fig_Y3}
\end{figure}

In this way we obtained a curve of $Y$ versus $N$ which reaches a minimum value
of $Y=0.590$ at $N=40$ and then increases with $N$. The largest system that
we could simulate ($N=100$)
looks to be still far from having attained the stationary
value, so we are unable to say if it agrees with the annealed prediction
$Y=2/3$, but it is very close to that value (figure \ref{fig_Y3}).

\begin{figure}
\centering
\epsfysize=15.0cm
\epsfxsize=10.0cm
\epsffile{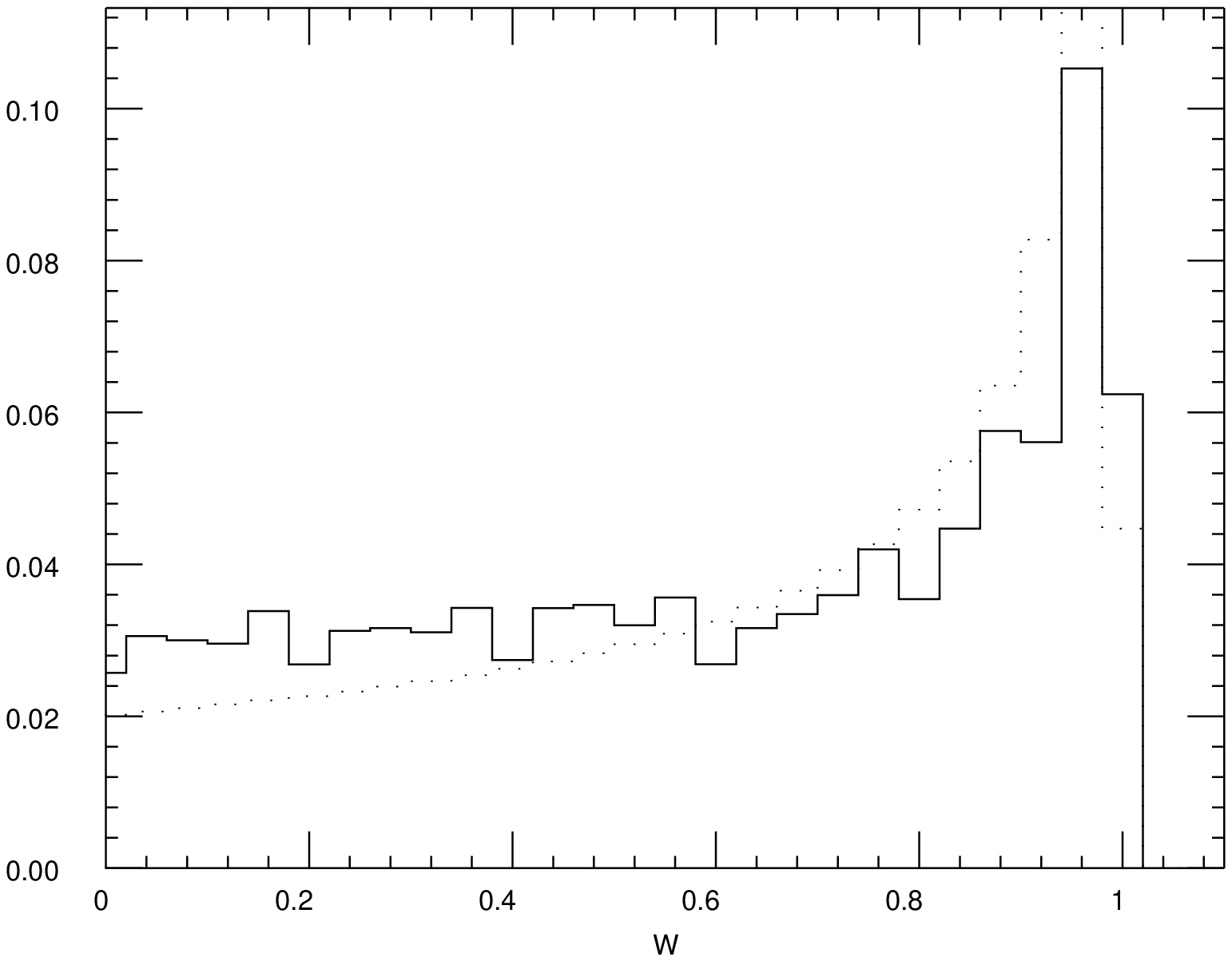}
\caption{\it Histogram of the probability to find a cycle of weigth $W$
in networks with $K=3$ and $N=50$, obtained simulating 1000 networks and
250 trajectories on each of them. The dotted line is the same thing,
computed in the framework of the annealed approximation.}
\label{pesi}
\end{figure}

We measured also the histogram of the probability to find a cycle of
weigth $W$. Figure \ref{pesi} compares the histogram obtained generating
at random 1000 networks with $K=3$ and $N=50$ and simulating 250 trajectories
on each of them with the one computed using (\ref{f(W)}), which is valid
in the infinite $N$ limit. The agreement, as it is possible to see, is not
so bad.

\section{Average cycle number}

In this section we aim to compute the average number of cycles in a network,
a quantity introduced by Kauffman to represent, in his biological metaphor, the
typical number of possible cell types for a genome of given length.
This quantity is difficult to observe in the simulations, because one must
consider also cycles of very small weight, and has been computed only for the
Random Map (infinite $K$, $\r=1/2$; \cite{DF2}).

For this purpose, we have to compute the distribution of
the overlap between a configuration at time $t$ and the initial configuration,
$P_N(0,t; q)$, which plays the role of the initial distribution for the
stochastic process defined by the master equation (\ref{master}).

It is easy to generalize the annealed approximation
in order to get the transition probability of the joint distribution of the
overlaps $q(0,t)$ and $q(t-1,t)$: every gene remains at time $t+1$ in the same
state it was at the previous step with probability $\g\left(q(t-1,t)\right)$,
where $\g(x)$ is defined in the previous section (\ref{gamma}).

If the genes
unchanged are $Nz$ of the genes whose state was the same in the initial
configuration and $N(q(t,t+1)-z)$ of the other genes, the overlap $q(0,t+1)$ is
given by $1-q(0,t)-q(t,t+1)+2z$. One then obtains the transition probability
\bea
&P\left\{q(0,t+1)=x_0,q(t,t+1)=x_1\mid q(0,t)=q_0,q(t-1,t)=q_1\right\}=
\hspace{2 cm}\nonumber\\
&\hspace{3cm}{Nx_0\choose Nz}{N(1-x_0)\choose N(x_1-z)}
\left(\g(q_1)\right)^{Nx_1}\left(1-\g(q_1)\right)^{N(1-x_1)},\eea
with
\be z={1\over 2}\left(x_1+x_0+q_0-1\right), \ee
and with the bounds
\be \left|1-(x_0+x_1)\right|\leq q_o\leq 1-\left|x_0-x_1\right|. \ee

At time $t=1$, the variables $q(0,t)$ and $q(t-1,t)$ coincide and their
distribution is a binomial one, but the correlations disappear after 2 time
steps, at least in the Gaussian approximation,
and the stationary joint distribution factorizes into the stationary
distribution of $q(t-1,t)$, that we met in the previous section, and a
binomial distribution of mean value 1/2. The most likely value of $q(0,t)$
is always 1/2, even before that the stationary distribution has been reached,
and the exponential part of the closing probability $\pi_N(0,t)$ (the
probability to have $q(0,t)=1$) is always $1/2^N$, but the prefactor is
different from 1, in the initial steps. As usual, this can be seen putting
the distribution in the exponential form and performing the integral with the
saddle point method.

The average number of cycles of length $l$ is related to the
probability $P(l)$ to find a configuration belonging to a cycle of length $l$
by the obvious formula
\be\bar n_l={2^N\over l}P\{L=l,T=0\}.\ee

In the framework of the annealed approximation,
using the results of section 4 about period and transient distribution
and the fact that $\pi_N(0,l)= 1/2^N c_l$ one finds, in the chaotic phase,
\be \bar n_l={c_l\over l}\exp\left(-{1\over 2}{l^2\over\t^2}\right).\ee

Summing over $l$ one gets the average number of cycles in a network.
Apart from systems at the critical point, $c_l$ is a quantity of order 1,
and tends to 1 when $l$ grows; so, substituting $c_l$ by 1 and transforming
the sum into an integral, one finds, asymptotically in $\t$, the following
behaviour:
\be \sum_{l=1}^{2^N}\bar n_l\approx \log\t={1\over 2}\a N,\ee
where $\a$ is the exponent of the stationary closing probability introduced
in the previous section.

This formula is valid only in the chaotic phase: in the
frozen phase, where $\a$ cancels, we are not allowed to put $c_l$ to 1.

For instance, for $l=2$ one finds
\be \pi_N(0,2)={1\over 2^N}\int {\rm e}^{-Nf(x)} (x(1-x))^{-1/2} {dx\over
\sqrt{2\pi/N}}, \ee
where
\be f(x)=x\log (x)+(1-x)\log (1-x)-x\log (\g(x))-(1-x)\log (1-\g(x)). \ee
($\g(x)$ is defined in (\ref{gamma})).
The function $f(x)$ is concentrated around the same point $Q^*$ around
which it is concentrated the stationary distribution of the overlap, $i.e.$
the solution of the equation $Q^*=\g(Q^*)$. The saddle point method with
Gaussian corrections yields the result
\be \pi_N(0,2)={1\over 2^N}\left(1-\g'(Q^*)\right)^{-1}, \ee
but at the critical point $\g'(Q^*)$ is equal to 1, so one has to go beyond
Gaussian corrections and finds
\be \pi_N(0,2)\propto {1\over 2^N} N^{1/6}. \ee

\section{Numerical tests of the Annealed Approximation}
All simulations reported in this section are referred to Kauffman Networks
with unbiased response functions, {\it i.e.} the parameter $\r$ is fixed at
the value $0.5$, and we will no longer mention it.

Our numerical analysis of the overlap distribution confirms the validity
of the annealed approximation in some range of the time variables
and reveals deviations from it in some other range,
so that the period distribution turns out to be different from the one
predicted by the annealed approximation, while the time-scale of the period
and the first moment of the distribution of the weights are in good agreement
with our computation.

\subsection{Overlap distribution}

We measured the overlap $q(t,t+l)$ on the subset of the trajectories not yet
closed at time $t+l-1$, in order to sample with the conditional probability
defined in section 2. In this way we also eliminate the
"deterministic noise" due to the periodicity of the orbits.

For each value of the parameters, we generated 20'000 sample networks and on
each of them we simulated a trajectory measuring the overlap
$q\left(C(t),C(t+l)\right)$ till the time when the trajectory closed.
So we obtained the histogram of the overlap $q(t,t+l)$ for chosen
values of the temporal distance $l$ and $t$ large enough to suppose that
the distribution is stationary.

{}From the annealed approximation we expect the stationary distribution not to
depend on $l$, but in fact we could observe different behaviours for different
$l$ values. So the data analysis is rather complex since for every point
$(K,\r)$ in parameter space three varables, $N$, $t$ and $l$, are involved
and different limits, {\it e.g.} $N$ large at fixed $t$ and $l$ or $t$ and $l$
large at fixed $N$, may give different behaviours.

We are interested to the exponential part of the stationary distribution,
and we estimate it through the formula

\be \a_l(q)\approx {1\over N}\left(-{1\over 2}\log N-\log P_{Nl}(q)\right),
\label{espo}\ee
in which the factor $1/\sqrt N$ coming from the Stirling expansion of the
binomial coefficient has been taken into account (actually, we use this formula
only for $0.1\leq q\leq 0.9$, while for $q=0$ and 1 we don't subtract the term
with $\log N$ and we interpolate linearly for intermediate values of $q$).

The function of $q$ so obtained is compared with the function $\a(q)$
numerically computed in the annealed approximation (equation \ref{map}).

Most of our simulations have been performed on systems with connectivity $K=3$.

\begin{figure}
\centering
\epsfysize=17.6cm
\epsfxsize=11.0cm
\epsffile{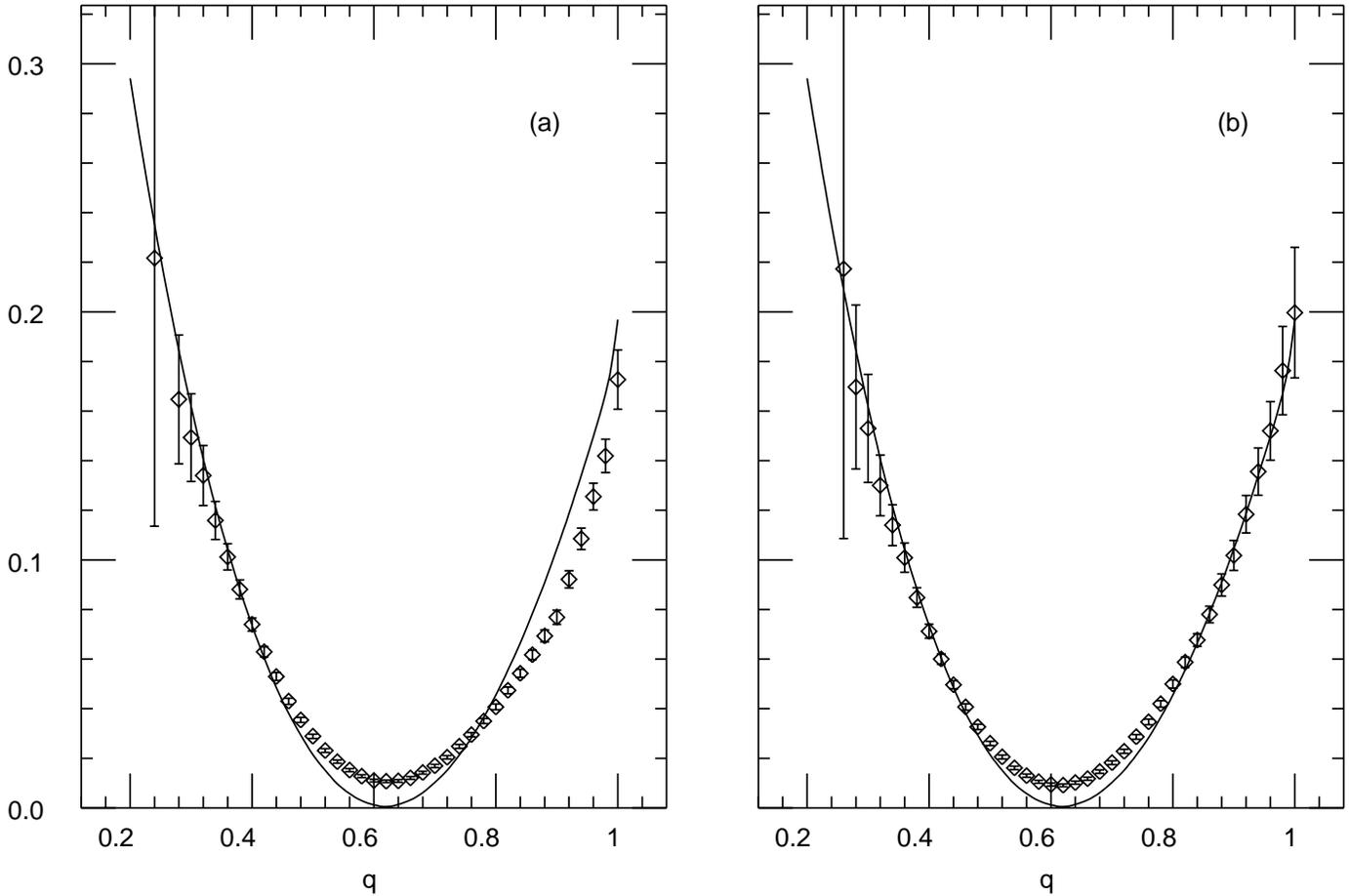}
\caption{\it The function $\a_l(q)$, which is the exponential part of the
distribution of the overlap $q(t,t+l)$. The solid line is the annealed
prediction, the points plotted come from the simulation of $20000$ networks
with $K=3$ and $N=50$. The distribution is obtained considering different
values of $t$ between 50 and 80.
(a): $l=2$; (b): $l=62$ (in this case the agreement for $q$ close to 1 is much
better). The condiction that the trajectory must not be closed at time $t+l-1$
selects respectively 79 and 63 per cent of trajectories in the sample.}
\label{fig_al3}
\end{figure}

The agreement is very good when $l$, the temporal distance between the
configurations, is large (in fig. \ref{fig_al3} (b)
data are shown for $l=62$, in networks with $K=3$
and $N=50$), while the shape of the distribution is different, especially
close to $q=1$, for small values of $l$ (in fig. \ref{fig_al3} (a) we show
data for $l=2$, in  networks with $K=3$ and $N=50$).

\begin{figure}
\centering
\epsfysize=17.6cm
\epsfxsize=10.0cm
\epsffile{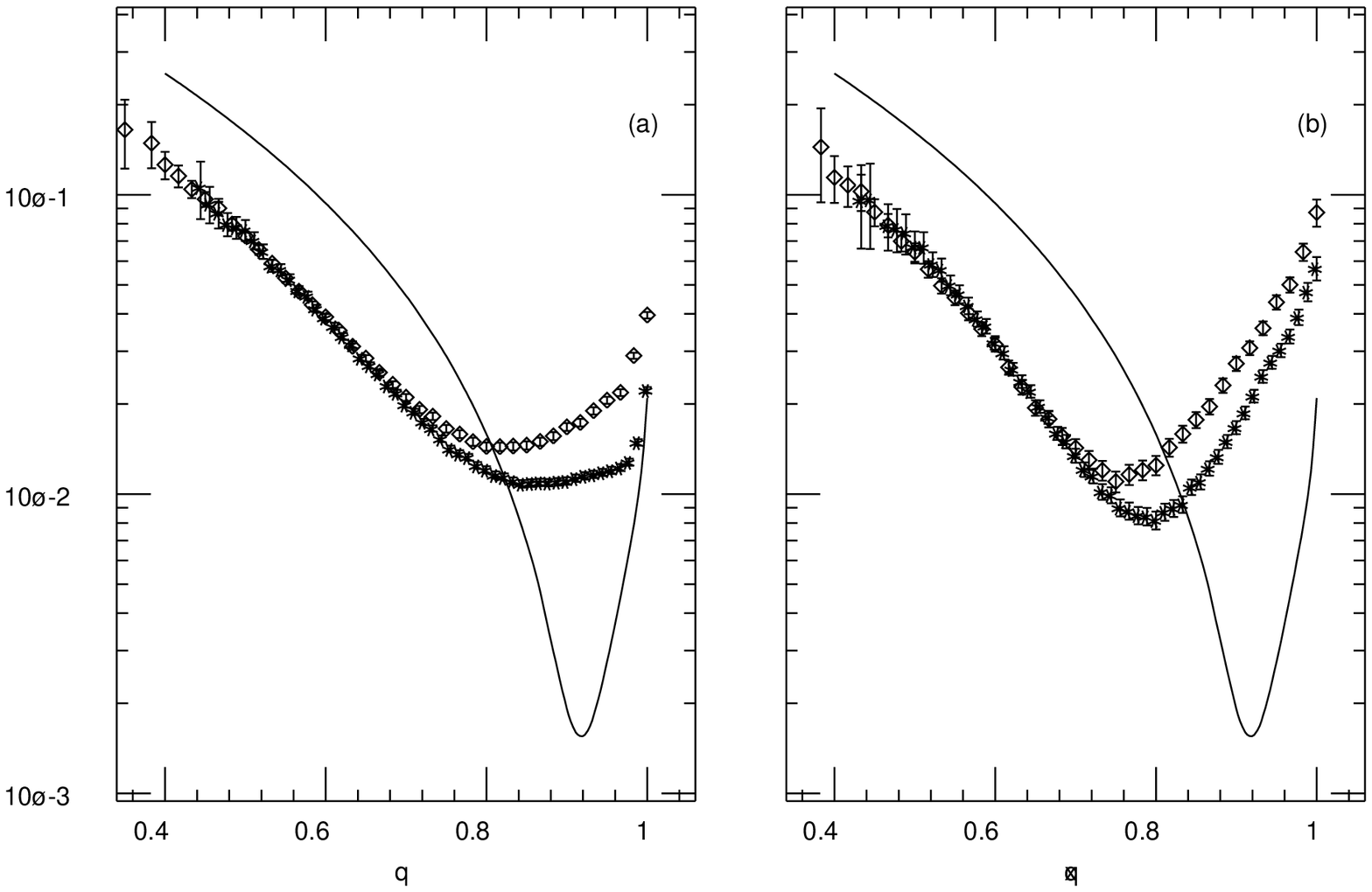}
\caption{\it The function $\a_l(q)$, which is the exponential part of the
distribution of the overlap $q(t,t+l)$. The solid line is the annealed
prediction for $K=2$ and $N=90$, the points plotted come from the simulation
of $40000$ networks with $K=2$, and $N$ respectively equal to $60$ (diamonds)
and $90$ (asterisks). The distribution is obtained considering different
values of $t$ between 12 and 20. The vertical scale is logarithmic.
(a): $l=2$; (b): $l=26$.
The condiction that the trajectory must not be closed at time $t+l-1$ selects
respectively 59 and $12.5$ per cent of trajectories in the sample for $N=60$
and
76 and 19 per cent for $N=90$.}
\label{bruco2}
\end{figure}

 In the frozen phase the agreement between the annealed approximation
and our data worsen. We simulated systems with $K=2$ and different
values of $N$. We show in figure \ref{bruco2} the function $\a(x)$
computed numerically within the annealed approximation and the one
measured in the simulation with $N=90$ using formula (\ref{espo})
(in this case, we subtracted the log term also for $q=1$, because we expect
$P(1)$ to be proportional to $1/\sqrt N$).


The agreement between data and computation is again better when the
temporal distance between the configurations compared is large, but the
real distribution is always much broader than the annealed one and the small
overlap values are much more likely in the simulation than in the annealed
approximation.

\vspace{0.5 cm}
Then we tried to see if the overlap between configurations on the limit
cycles is statistically different from the overlap between transient
configurations. For such purpose we computed the function $\a_l(q)$
imposing two kind of conditions: on one hand,
that both configurations be on the cycle, on the other
that both configurations be transient.

In the chaotic phase (we report data with $K=3$ and $N=50$)
the functions so obtained fit
within one or two standard deviations (which are large, because
the selection of trajectories is very severe) to the annealed prediction,
for the same values of parameters for which the overall distribution gives
a good function $\a(q)$, and so we don't see differences in the overlap
distribution between pairs of transient configurations and
pairs of configurations in the limit cycles.

 The situation is different in the frozen phase where one can see that
the most probable value of the overlap between transient configurations is
quite lower than the most probable overlap between configurations of a cycle.
\begin{figure}
\centering
\epsfysize=17.6cm
\epsfxsize=10.0cm
\epsffile{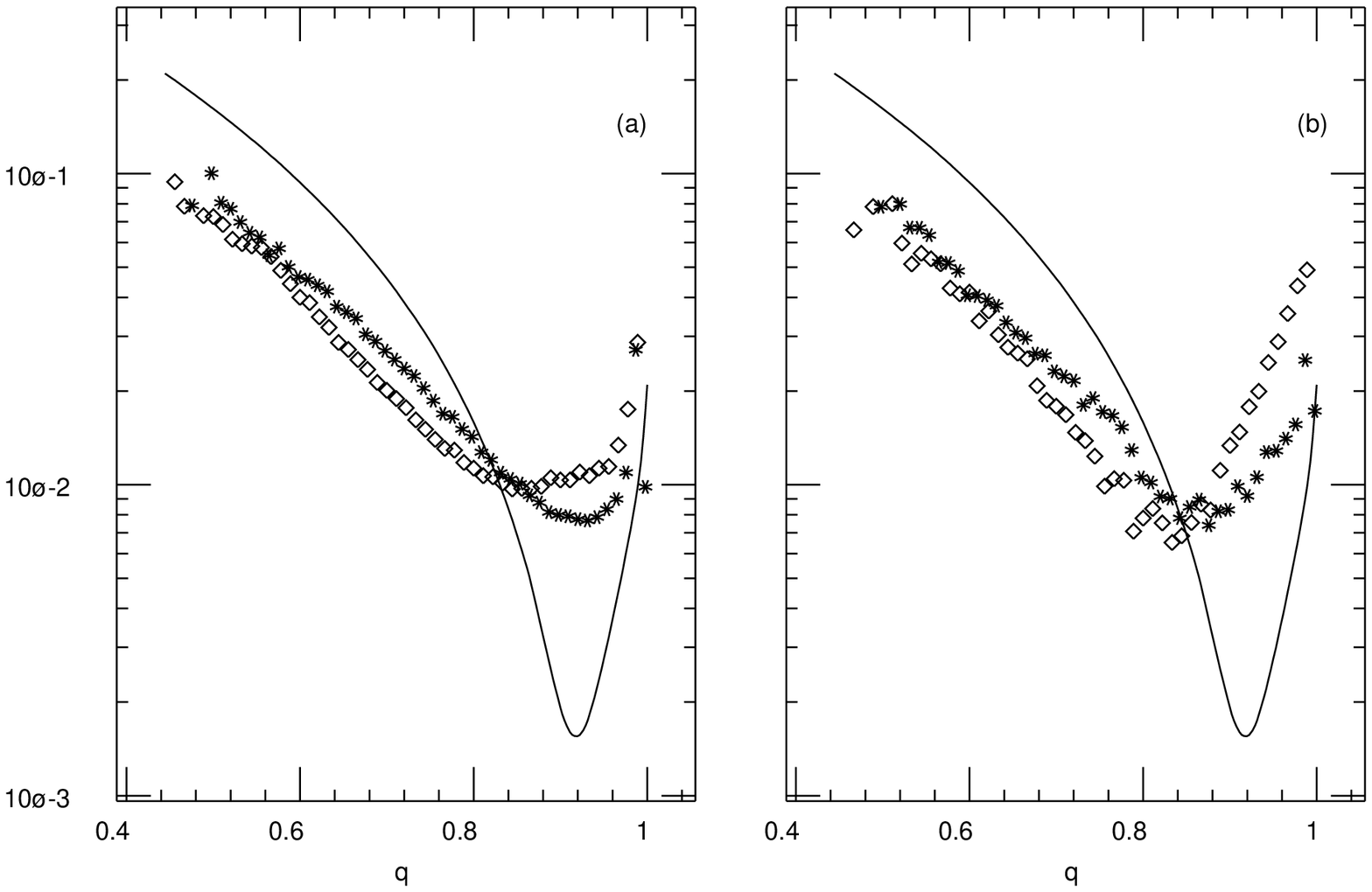}
\caption{\it The function $\a_l(q)$ again. Here $K=2$ and $N=90$.
The solid line is the annealed prediction, the points plotted come from the
simulation of $40000$ networks.
The distribution of the overlap $q(t,t+l)$ is obtained considering different
values of $t$ between 12 and 20. The vertical scale is logarithmic.
The two graphs concern $l=2$ (a) and $l=14$ (b) respectively.
The asterisks are refered to the overlap of two configurations in a limit
cycle,
the diamonds to the overlap between transient configurations.}
\label{brucoc2}
\end{figure}
\begin{figure}
\centering
\epsfysize=17.6cm
\epsfxsize=10.0cm
\epsffile{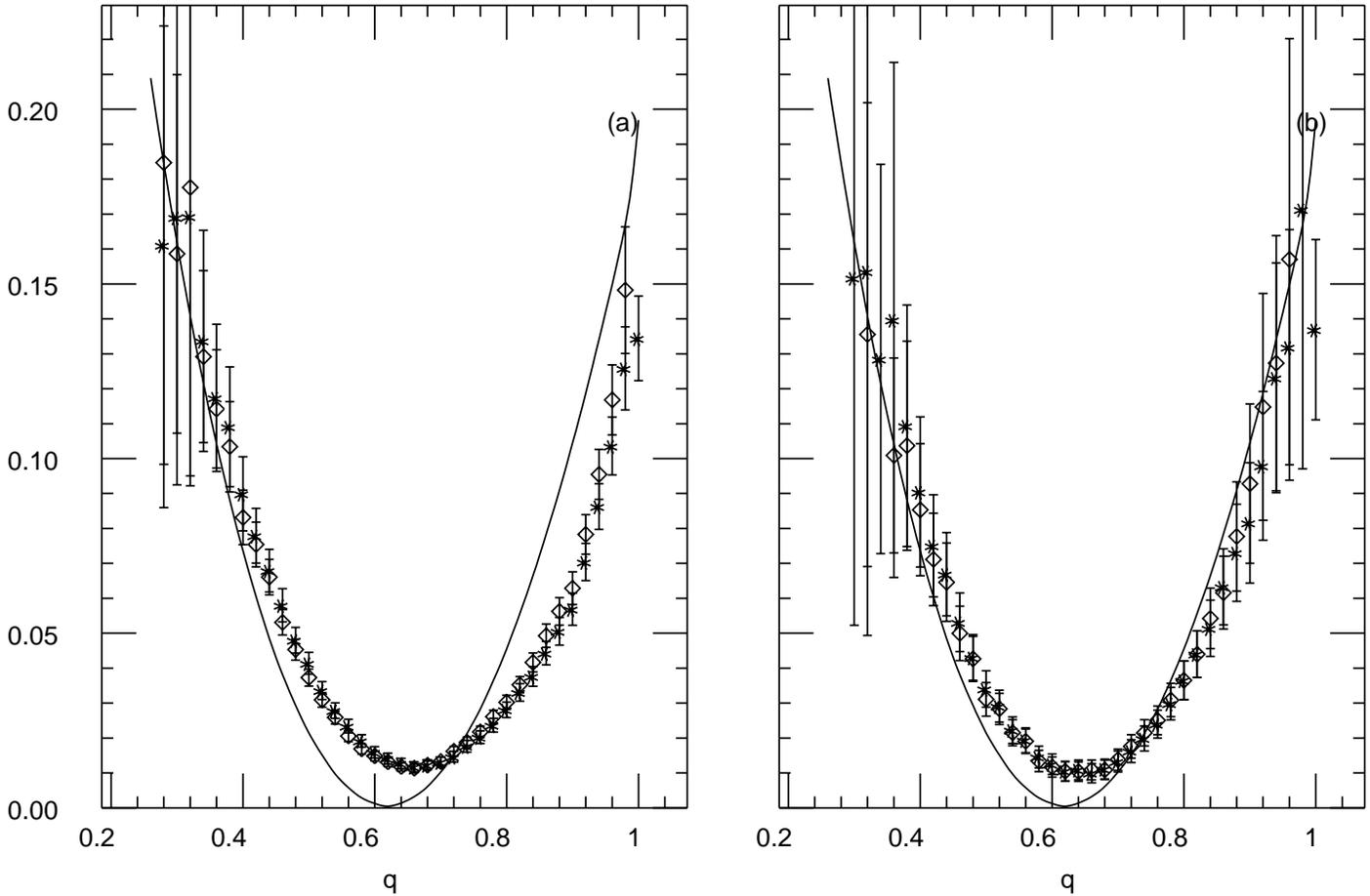}
\caption{\it The function $\a_l(q)$ again. Here $K=3$ and $N=50$.
The solid line is the annealed prediction, the points plotted come from the
simulation of $20000$ networks.
The distribution of the overlap $q(t,t+l)$ is obtained considering different
values of $t$ between 50 and 80.
The two figures are concerned with $l=2$ (a) and $l=62$ (b) respectively.
The asterisks are refered to the overlap of two configurations in a limit
cycle,
the diamonds to the overlap between transient configurations.}
\label{brucoc3}
\end{figure}

\subsection{Measure of the closing probabilities}
In section 2 we defined the closing probability $\pi_N(t,t+l)$ as the
probability that the configurations at times $t$ and $t+l$ be equal, with the
condition that the trajectory is not yet closed at the previous time steps.

We studied the closing probability for $l$ fixed as a function of $t$,
and we saw that in the quenched model the overlap distribution doesn't go
to a stationary distribution as it is predicted by the annealed approximation.

For a fixed system size, the closing probability grows to a maximum value,
usually larger than the annealed stationary value (but this depends on $l$!)
and after it is always decreasing.

The rate of decrease appears to be independent on $l$, while the maximum value
attained is not: it is larger if $l$ is even and decreases with $l$ for a given
parity.

\begin{figure}
\centering
\epsfysize=17.6cm
\epsfxsize=10.0cm
\epsffile{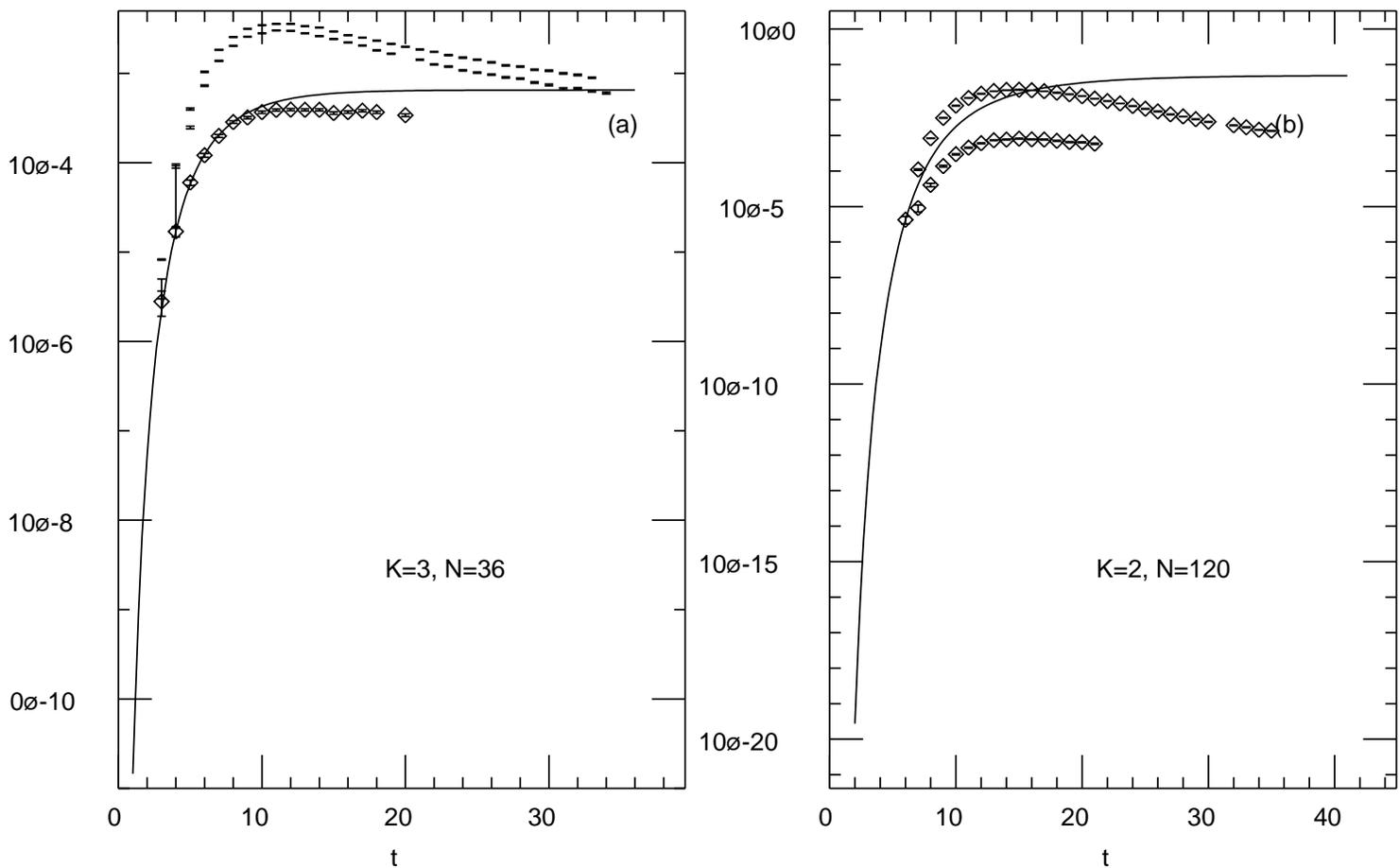}
\caption{\it Closing probability $\pi_N(t,t+l)$. In solid
it is reported the annealed value with $l=1$.
The ordinate scale is logarithmic.
(a): The system is in the chaotic phase, $K=3$ and $N=36$.
Data correspond to $l=1$ (middle),
$l=2$ (upper line: the closing probability is systematically higher) and to
$l=15$ (diamonds; the agreement with the annealed approximation is better).
(b): The system is at the critical point, $K=2$ and $N=120$. Data correspond
$l=2$ (upper line) and to $l=15$ (diamonds). The first point of the annealed
computation, $P=7.52\cdot 10^{-37}$, is not shown.}
\label{fig_paiann}
\end{figure}

In figure \ref{fig_paiann} we report data for $K=2$, which, in our case
$\r=0.5$, is the starting point of the frozen phase, and $K=3$ in the chaotic
phase.

The simulation results are compared to the annealed closing probability exactly
computed iterating equation (\ref{master}). The time
behaviour is at first in good agreement (especially for $l$ large), over a
fairly large range: in the networks with $K=2$ and $N=120$ the annealed and the
quenched grow together from $10^{-36}$ to $10^{-2}$. After, the annealed
remains constant while the quenched decreases.

On the other hand, when we compared to the annealed approximation the closing
probability measured at fixed times varying system size, we found out that
the agreement, at the beginning not very good (many standard deviations),
becomes very satisfactory taking larger systems.


Put into a graphic, the $N$ behaviour of the closing probability appears to be
an exponential decay with small corrections.


\vspace{0.3 cm}

We now study the integral closing probability, $\tilde\pi_N(t)$, which is the
sum over $t'$ of $\pi_N(t',t)$. As usual, we considered two values of $K$, 2
and 3, in and near to the frozen phase, and $N$ ranging from 60 to 180 for
$K=2$ and from 20 to 50 for $K=3$.
\begin{figure}
\centering
\epsfysize=15.0cm
\epsfxsize=10.0cm
\epsffile{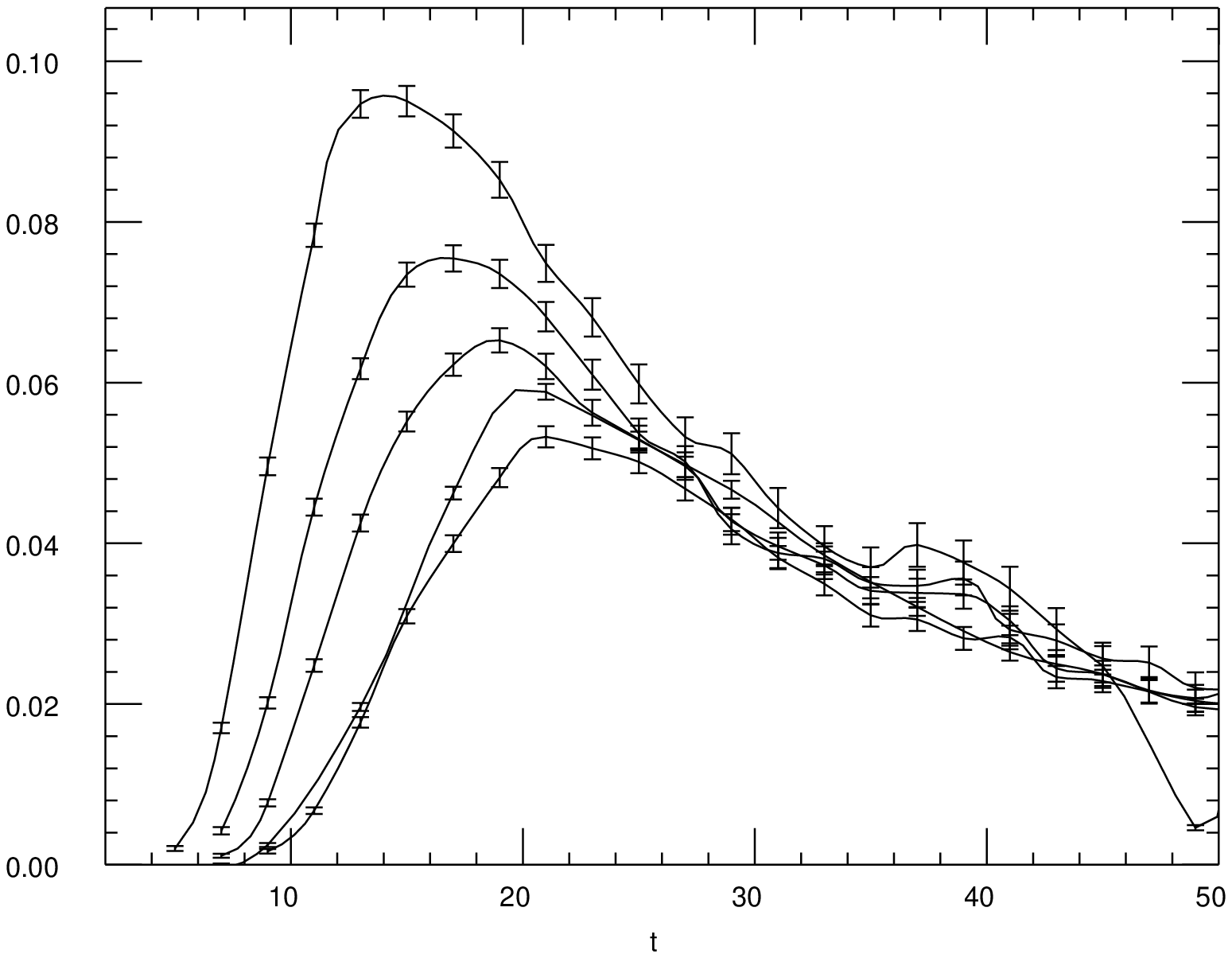}
\caption{\it Integral closing probability, $\tilde\pi_N(t)$ in the frozen
phase ($K=2$) for different system sizes: $N=60, 90, 120, 150, 180$.
$\tilde\pi_N$ reduces when $N$ grows.}
\label{fig_pai2}
\end{figure}

\begin{figure}
\centering
\epsfysize=15.0cm
\epsfxsize=10.0cm
\epsffile{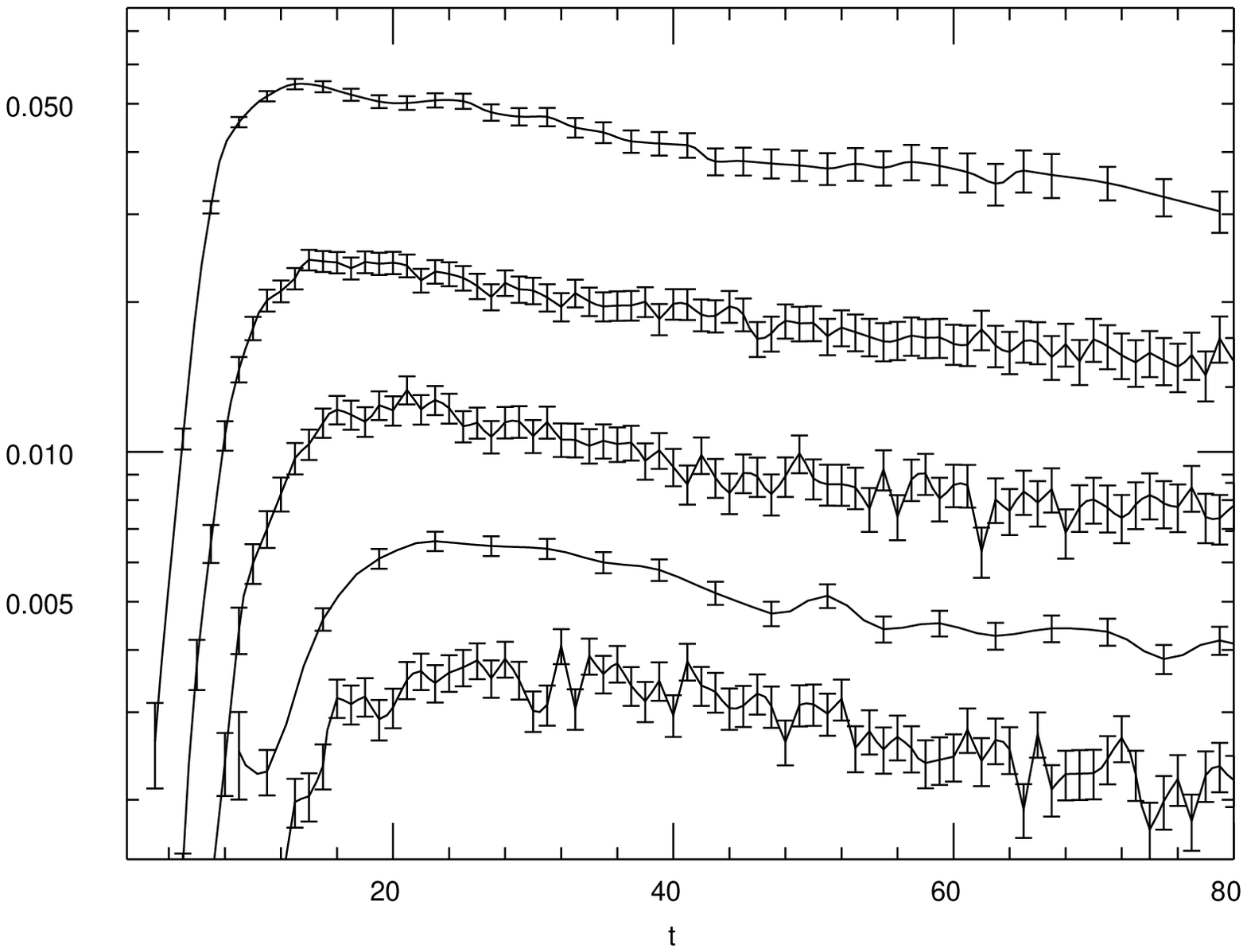}
\caption{\it Integral closing probability, $\tilde\pi_N(t)$ in the chaotic
phase ($K=3$) for different system sizes: $N=20, 30, 40, 50, 60$.
$\tilde\pi_N$ reduces when $N$ grows.}
\label{fig_pai3}
\end{figure}

Also $\tilde\pi_N(t)$
has a non monotone behaviour for $t$ not too large: it increases to
a maximum value and then start decreasing. Of course, it has to increase at
large times, because it has to fulfil the normalization condition
$\tilde\pi\left(2^N\right)=1$, but in our simulations
it keeps decreasing at all times accessible to measurement. In the framework
of the annealed approximation, we expect that it grows linearly with $t$.

There is nothing new in this behaviour, which is just a consequence of the
decrease in time of the closing probability, but it is the origin of other
interesting features of the quenched system.
 In a next subsection we shall try to give an interpretation to this funny
result, which is in contrast with the annealed approximation.


We show the temporal behaviour of $\pi_N(t)$ for different system sizes in the
chaotic phase (fig. \ref{fig_pai3}) and in the frozen one (fig.
\ref{fig_pai2}).
The two plots have qualitatively
the same shape, but in the frozen phase the integral closing probability
decreases very slowly with system size $N$, while in the chaotic phase it
decays exponentially.
\vspace{0.3 cm}
\begin{figure}
\centering
\epsfysize=17.6cm
\epsfxsize=10.0cm
\epsffile{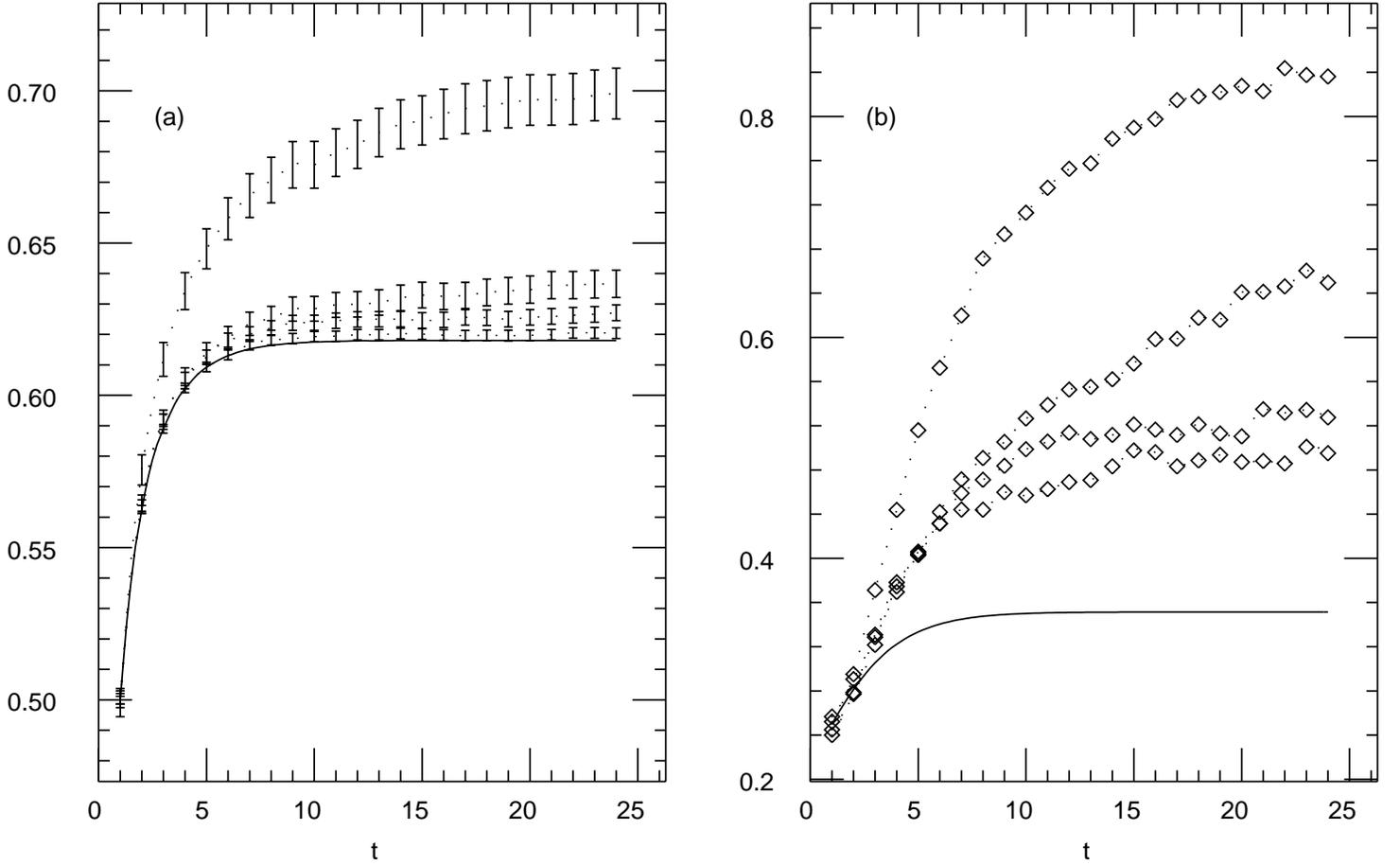}
\caption{\it (a): Average overlap between successive configurations
The solid line shows the annealed equation, $Q(t+1)={1\over 2}\left(1+Q(t)^K
\right)$.
(b) Variance of the overlap between successive configurations.
The solid line shows the annealed equation, $V(t+1)=Q(t+1)\left(1-Q(t+1)
\right)+\left({K\over 2}Q(t)^{K-1}\right)^2 V(t)$. The system is in the chaotic
phase, $K=3$; data are shown for $N=30$, 80, 200 and 400; the larger is $N$,
the lower is the curve.}
\label{fig_q3}
\end{figure}

We also measured the mean overlap between successive configurations,
$\bar q(t,t+1)$, without limiting the sample to the only trajectories not yet
closed. We considered three different connctivities in the chaotic phase
($K$=3,4 and 5) and varied $N$.

As a function of $t$, this quantity increases to a stationary value which
tends,
when $N$ grows, to the same value $Q^*$ computed within the annealed
approximation, but the (positive) corrections are large
even for system sizes of the order of hundreds; on the other hand, when the
temporal distance between the configurations compared is large, the corrections
are much smaller.

This measurement was done without imposing the opening condition. When we
measured the mean overlap between successive configurations, $\bar q(t,t+1)$,
on the subset of the trajectories not yet closed at time $t+l$, we noticed
that this quantity has a non monotone behaviour. It starts at $t=0$ with
the value $0.5$, then increases, reaches a maximum value at about the same time
at which the integral closing probability has its maximum value, and decreases
again. Thus, the behaviour of the mean overlap parallels that of the closing
probability.

\begin{figure}
\centering
\epsfysize=15.0cm
\epsfxsize=10.0cm
\epsffile{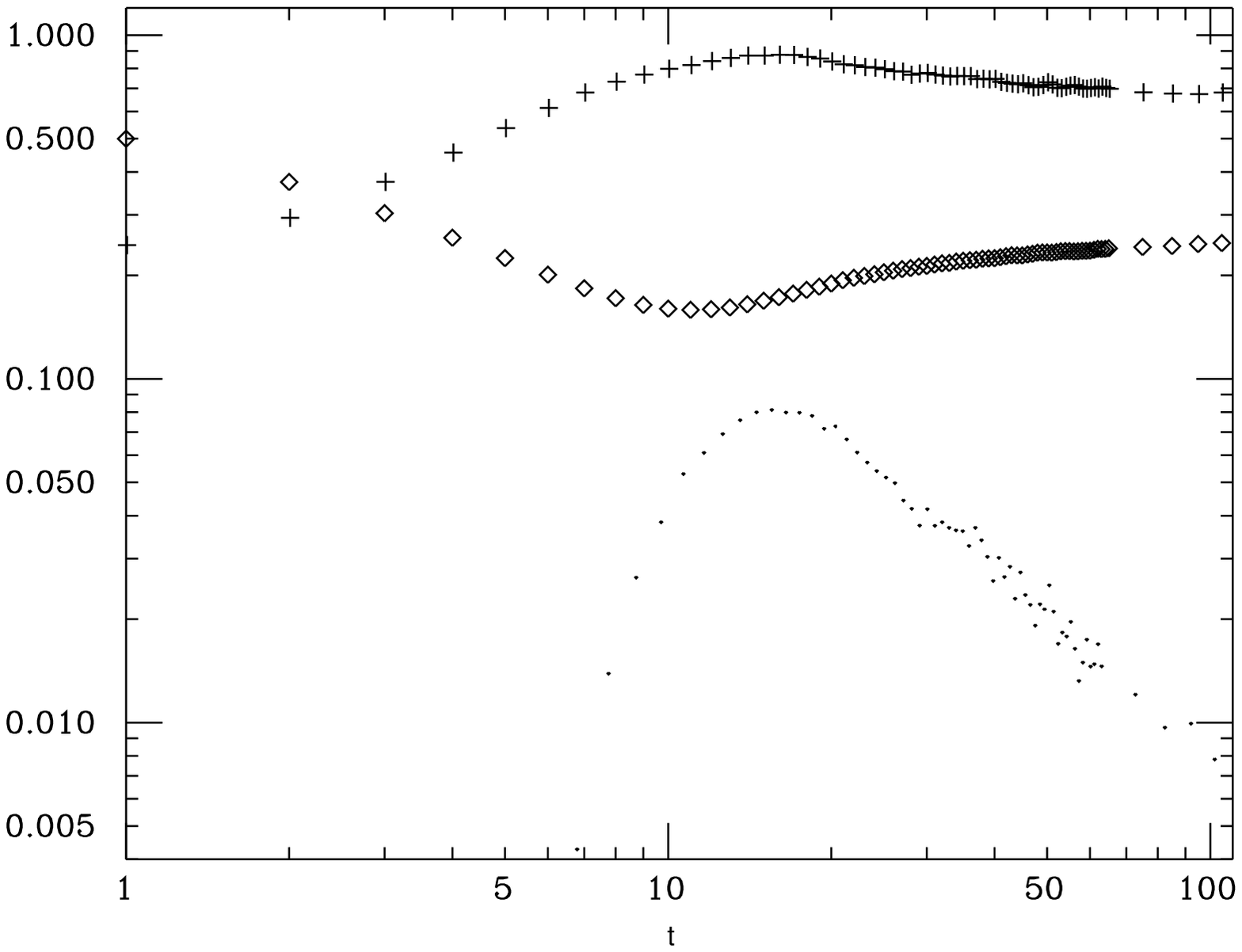}
\caption{\it Average ($\diamondsuit$) and variance (+) of the distance $d=1-q$
between successive configurations
on trajectories not yet closed at time $t-1$, as a
function of $t$. It is also shown the integral closing probability,
$\tilde\pi_N(t)$ ($\cdot$). $K=2$, $N=90$, $10000$ sample networks. }
\label{fig_qs}
\end{figure}

This behaviour indicates that the condition that the trajectory is not
yet closed at a certain time selects, from a point on,
smaller and smaller overlaps. Fig. \ref{fig_qs}
shows the mean and the variance of the distance $d=1-q$ of successors
on trajectories not yet closed and the integral closing probability for $K=2$
and $N=90$. The effect is more evident in the frozen phase, but it is still
present in the chaotic phase for low values of $K$.

\subsection{Properties of the cycles}
Our simulations confirm well the predictions of the annealed approximation
concerning the time scale of the period and of the transient distribution.

In the chaotic phase these quantities
have an exponential behaviour, $\exp(\a N/2)$. We simulated systems
with $K$ ranging from 3 to 6 and several values of $N$ (from 10 to 100 for
$K=3$, from 16 to 40 for $K=4$, from 16 to 31 for $K=5$ and from 6 to 28 for
$K=6$), generating at least
10000 samples of each network and few configurations on each network (except
for the case $K=3$, where we generated hundreds of initial configurations
in order to compare the variance of cycle length on a given network with the
variance between different networks).

We fixed a maximum time
for the simulation, exponentially increasing with $N$; in such a way, we
underestimated the average cycle length because some
trajectories where left before they completed their limit cycle; they were
about one percent in the worst cases, but we think that this does not affect
very much the determination of the exponents of the average period.

\begin{figure}
\centering
\epsfysize=15.0cm
\epsfxsize=10.0cm
\epsffile{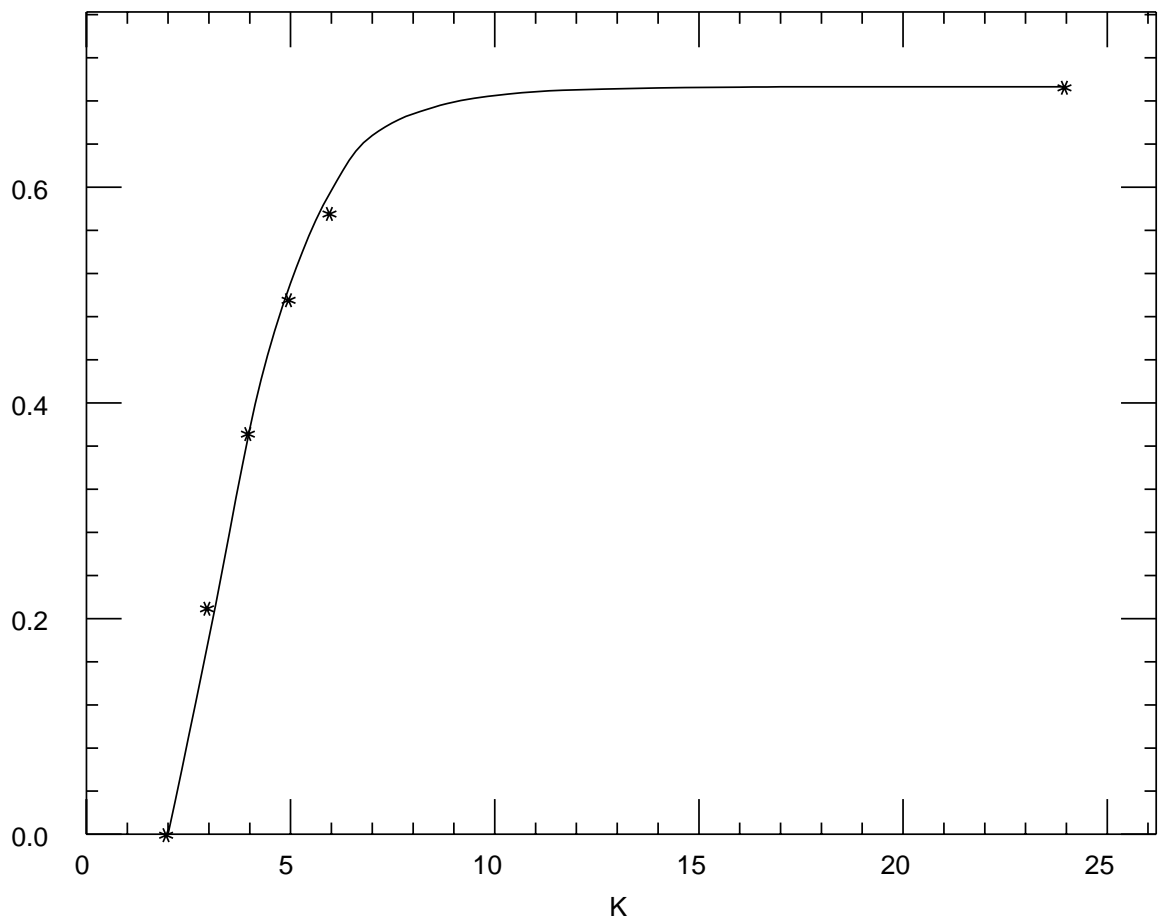}
\caption{\it The function $\a(K)$, where $\a$ is
twice the exponent of the characteristic time-scale of period and
transient distribution, $\t=\exp\left(\a N/2\right)$,
for $\r=0.5$, as obtained
from the annealed approximation. The asterisks are obtained from a fit of
simulational data; the last one is the theoretical result for the Random Map.}
\label{fig_alp}
\end{figure}

In fig.\ref{fig_alp} we plot the exponents $\a$ of the time scale of
the system, defined by $\t=\exp\left(\a N/2\right)$, and compare it
with the annealed prediction, numerically computed.

It is possible to measure in the simulations four different exponents:
$\a_L$ (twice the mean period exponent), $\a_V$ (the exponent of the
variance of the period), $\a_T$ (twice the mean transient exponent),
and $\a_P$, the exponent of the time-scale of the period distribution
(defined by fitting the probability to find a period larger than $t$ with the
function $\exp\left(-(t/\t)^\b\right)$).

These exponents were obtained through a fit of data with different $N$ values.
Their values lie very close whithin a few percent. In the figure there are not
error bars, because the fits of our data are affected by finite size effects
whose importance is difficult to estimate, but the agreement is quite good,
even with not very large systems.

The largest deviation was found for the case $K=3$,
where the annealed approximation gives $\a=0.182$ while our simulations
give $\a_L=0.210$, $\a_T=0.204$, $\a_P=0.21$ and $\a_V=0.206$ for the average
variance in a given network and $0.205$ for the variance
between different networks.

\begin{figure}
\centering
\epsfysize=15.0cm
\epsfxsize=10.0cm
\epsffile{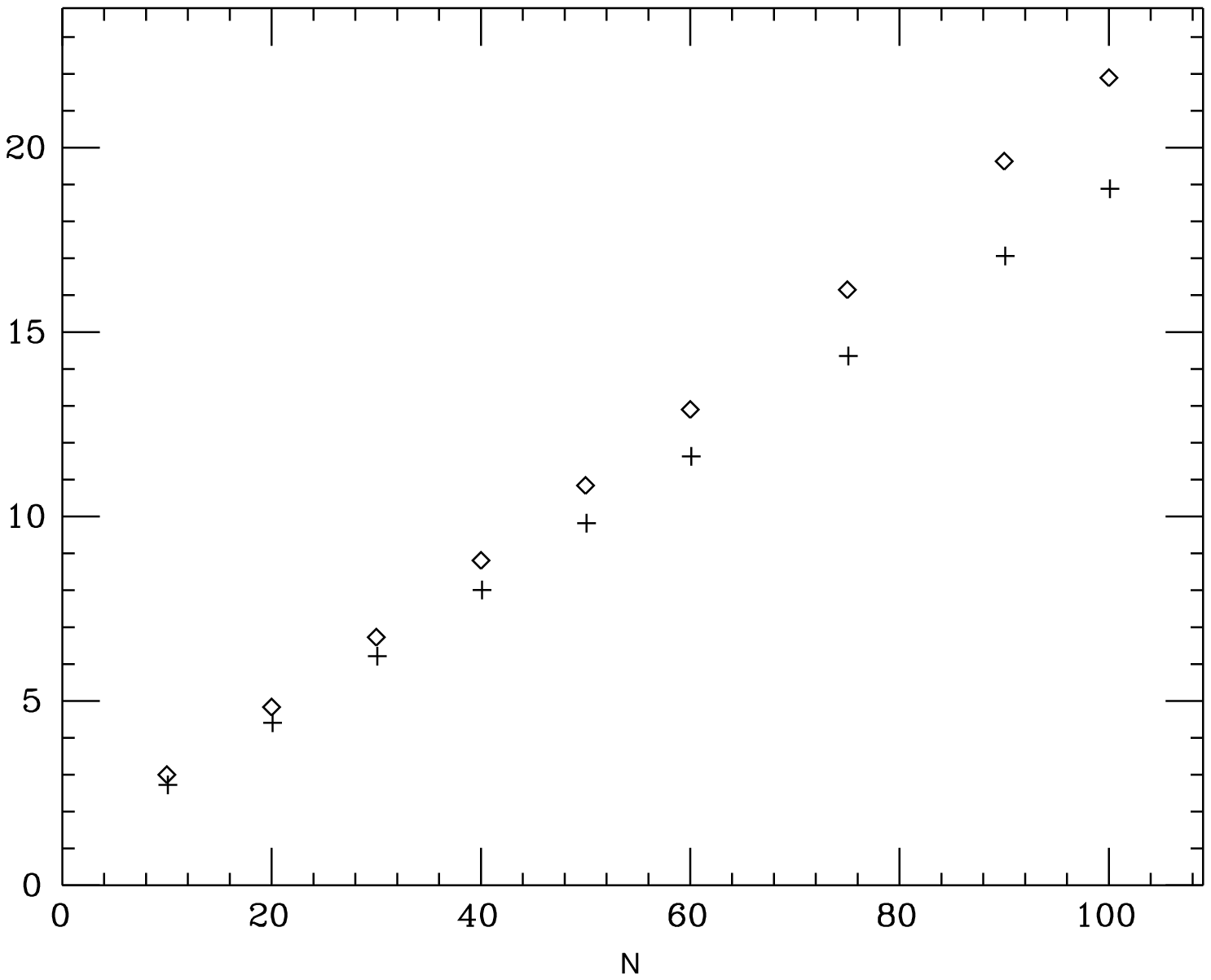}
\caption{\it ($\diamondsuit$) : twice the logarithm of the average period in
networks with $K=3$ and $N$ nodes (simulation); (+) : the annealed coefficient
$\a$ computed in the saddle point approximation for a system of $N$ nodes,
times $N$.}
\label{taulog}
\end{figure}

In this case, we performed the annealed computation with the map (\ref{map})
and the same values of $N$ used in the simulations, in order to obtain the
corrections due to the discretization, whose effect is to increase
$\a$. In figure \ref{taulog} we compare twice the logarithm of the average
period (diamonds) with $N\a(N)$, where $\a(N)$ is computed with the annealed
approximation. The discrepancies now are reduced respect to the asymptotic
$\a$ value, but the simulation data still seem to be systematically larger
and the angular coefficients are different.

\vspace{0.5 cm}
In the simulations with $K=3$ we measured also $F_N(t)$, defined in the second
section as the probability that a random chosen trajectory is not yet closed
at time $t$, and related to the closing probability by formula (\ref{F}).
According to the annealed approximation, one should find $\log\left(F_N(t)
\right)\propto t^2$, because the closing probability reaches a constant value.
In fact one finds that this quantity is linear or less than linear in $t$, so
that $F_N(t)$ has a stretched exponential behaviour.

This feature is not surprising because we have already seen that the
integral closing probabiity is a decreasing function of $t$ from a point on,
and $\log\left(F_N(t)\right)$ is just its integral, but it is striking
that the exponent of $t$ is a slightly decreasing function of $N$.

This last observation was made looking at the period
distribution which is a quantity much easier to compute: to measure $F_N(t)$
one has to keep in memory the whole trajectory, while to measure the period
one needs only a reference configuration.

The distribution of cycle length $L$ and transient time $T$ are simply related
to $F_N(t)$:
\be {\rm Prob}\left\{T=t,L=l\right\}=F_N(t+l)\pi_N(t,t+l). \ee

 For $K=3$,
the probability to find a cycle of period greater than $t$ decays as a
stretched exponential, $\exp\left(-(t/\t)^\b\right)$; the time scale $\t$
grows exponentially with $N$, with the same exponent $\a_L$ of the average
cycle length; the decay exponent $\b$ goes down with $N$,
even if very slowly: we found an exponent of $0.91$ for $N=10$ and $0.39$
for $N=100$. This fact is in contradiction with the results of the annealed
approximation, which predicts $\b=2$, but it is a direct consequence of the
decrasing with time of the closing probability.

{}From our simulations on systems with $K=2$ and $N$ ranging from 30 to 90
and $K=3$ and $N$ ranging from 10 to 100 it appears that
the histogram of transient time has a peak corresponding to the time $t$
where the closing probability reaches its maximum value, and then decreases
as a stretched exponential; and the cycle length histogram decreases with the
same law.

\begin{figure}
\centering
\epsfysize=15.0cm
\epsfxsize=10.0cm
\epsffile{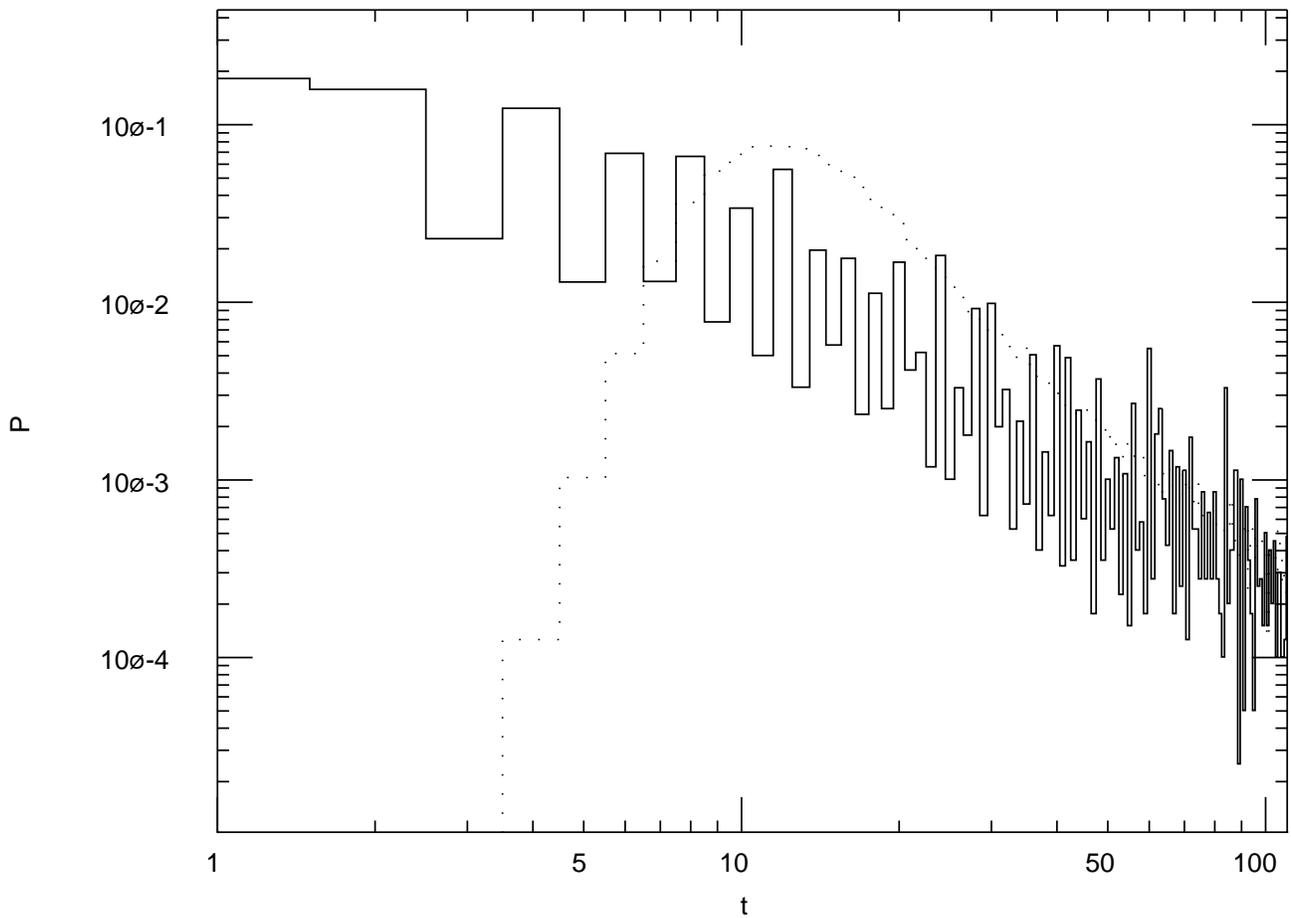}
\caption{\it Histogram of cycle (solid line) and transient (dots) length
for a system in the frozen phase, $K=2$ and $N=120$. $10000$ sample networks
were generated.}
\label{fig_pertrans}
\end{figure}

In addition, very short periods
are much more common respect to the others than one would expect on the basis
of the annealed approximation (see figure \ref{fig_pertrans}).

Another unexpected feature of the period distribution is the fact that cycles
of even lengths are more likely than odd ones, as one can already see by the
fact that the closing probability is higher for even periods.

We have seen in this fact a
clue of a non trivial distribution of local periodes, say the period of a
single
element in a network. We have obtained in our simulations an indirect
confirmation of this guess. We will present these data in a forecoming paper.

\begin{figure}
\centering
\epsfysize=15.0cm
\epsfxsize=10.0cm
\epsffile{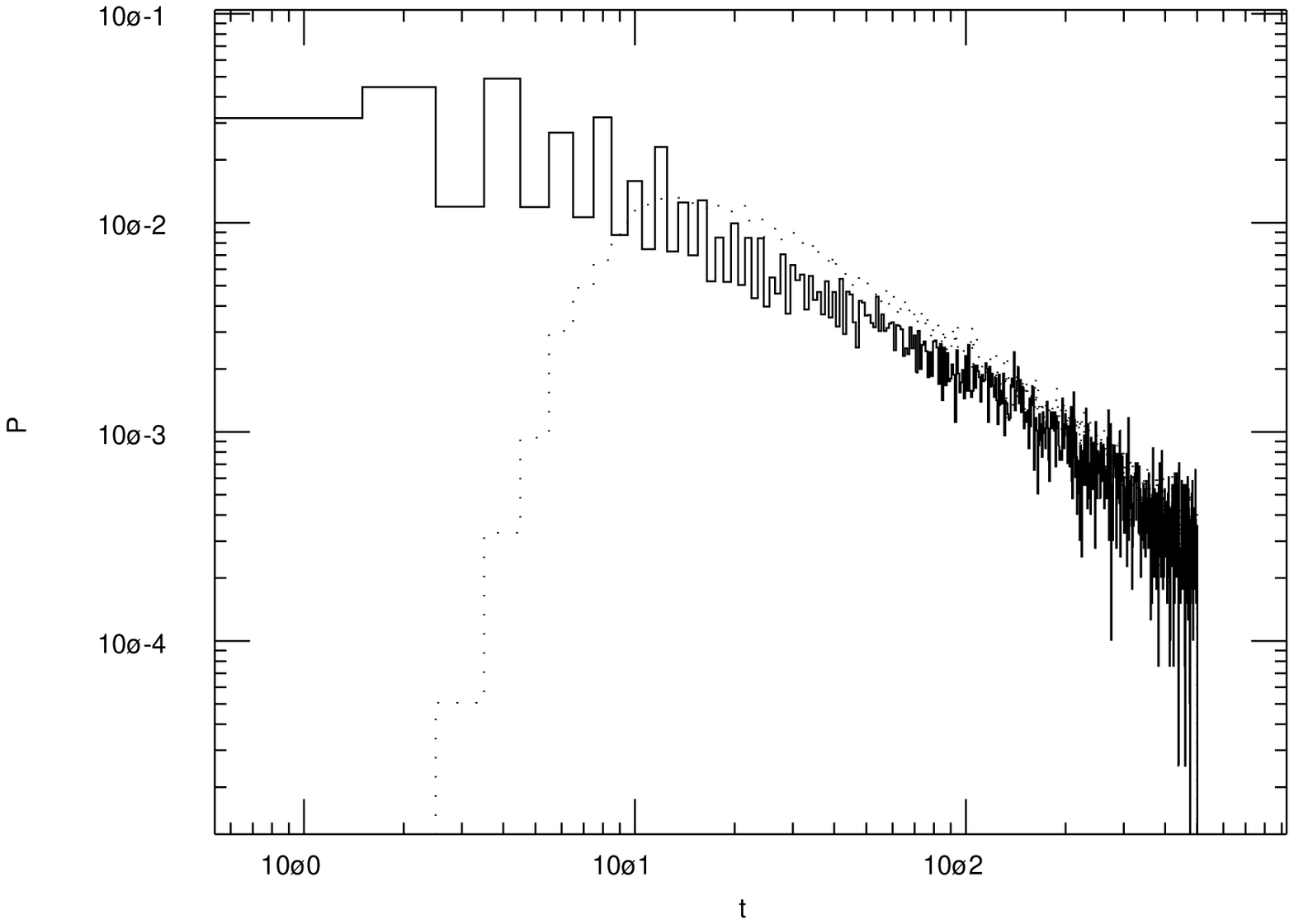}
\caption{\it Histogram of cycle (solid line) and transient (dots) length
for a system in the chaotic phase, $K=3$ and $N=50$. $10000$ sample networks
were generated.}
\label{fig_pertrans3}
\end{figure}

\section{Discussion}
In this work we used a stochastic scheme, generalizing the analysis by Derrida
and Flyvbjerg of the Random Map Model to finite values of $K$, in order to
compute the distributions of cycle lengths, transient times and attraction
basins weigths in Kauffman Networks.
The definition of the closing probabilities is crucial in this scheme.

The annealed approximation, introduced by Derrida and Pomeau, has revealed
to be a good tool for the computation of the closing probabilities and the
approximate solution of the model.

There are deviations from the approximation, which are larger close to the
frozen phase, but the mean quantities that we computed are in good agreement
with our numerical results. The annealed approximation works better in the
chaotic phase, but it allows also to predict an universal behaviour of cycle
lengths along the critical line. We have in plan to check this point.

We haven't, till now, tried to justify the approximation used.
The annealed approximation allows to compute the overlap distribution using
only a minimal information about the past evolution of the system and
neglecting everything else, so we think that a loss of memory of the history of
the system is needed for its validity.
This fact is consistent with our numerical results, which show that the overlap
of two configurations very far apart along the trajectory is in some sense
"more annealed'' than the overlap of two configurations temporally closer.
How this loss of memory comes about and what are its limitations would
be an interesting question to address to understand in a more analytical way
dynamical systems with quenched disorder.

As we saw in the previous section, the main discrepancies between
our simulations and the annealed approximation arise from the fact that the
closing probabilities ultimately decrease in time.

This fact can be interpreted as a consequence of the definition of the
closing probabilities as conditional probabilities: the requirement to consider
only trajectories not yet closed selects, from a time on,
trajectories where the typical overlaps are smaller and smaller.

We did not impose the opening condition in our computation, so we think that
the best way to improve the annealed approximation would be to take into
account the opening of the trajectory in the computation.
This purpose requires to keep memory of
much more information about the past history of the trajectory than the simple
one that we used in doing the annealed computation. We are planning to try
to do this in a next work.

\section{Aknowledgments}
We thank Bernard Derrida for a stimulating discussion.
Ugo Bastolla thanks Luca Peliti and Henrik Flyvbjerg
for interesting discussions and for reading the manuscript.

\end{document}